\documentclass{PoS}

\usepackage[authoryear,square]{natbib}
\let\ifpdf\relax
\bibpunct{(}{)}{;}{a}{}{,}

\title{Exploring AGN Activity over Cosmic Time with the SKA}

\ShortTitle{Radio AGN Activity over Cosmic Time}

\author{
\speaker{Vernesa Smol\v{c}i\'{c}}$^1$, 
Paolo Padovani$^2$,
Jacinta Delhaize$^1$,
Isabella Prandoni$^3$,
Nicholas Seymour$^4$,
Matt Jarvis$^{5,6}$,
Jose Afonso$^{7,8}$,
Manuela Magliocchetti$^9$
Minh~Huynh$^{10}$,
Mattia Vaccari$^{6}$,
Alexander Karim$^{11}$
\\ 
$^1$University of Zagreb, Physics Department, Bijeni\v{c}ka cesta 32, 10002 Zagreb, Croatia;
$^2$European Southern Observatory, Karl-Schwarzschild-Strasse 2, 85748 Garching b. M\"{u}nchen, Germany; 
$^3$INAF - Istituto di Radioastronomia, Bologna, Italy; 
$^4$International Centre for Radio Astronomy Research, Curtin University, Perth, Australia; 
$^5$ Astrophysics, University of Oxford, Keble Road, Oxford, OX1
       3RH, UK;
$^6$ Physics Department, University of the Western
       Cape, Bellville 7535, South Africa;
$^7$Instituto de Astrof\'{i}sica e Ci\^{e}ncias do Espa\c co, Universidade de Lisboa, OAL, Tapada da Ajuda, PT1349-018 Lisboa, Portugal;
$^8$Departamento de F\'{i}sica, Faculdade de Ci\^{e}ncias, Universidade de Lisboa, Edif\'{i}cio C8, Campo Grande, PT1749-016 Lisbon, Portugal; 
$^9$INAF-IAPS, Via Fosso del Cavaliere 100, I-00133 Roma, Italy;
$^{10}$International Center for Radio Astronomy Research, M468, University of Western Australia, Crawley, WA 6009, Australia;
$^{11}$Argelander Institut for Astronomy, Auf dem H\"{u}gel 71, Bonn D-53121, Germany
\\
E-mail: \email{vs@phy.hr}
}

\abstract{

In this Chapter we  present the motivation for undertaking both a wide and deep survey with the SKA in the context of studying AGN activity across cosmic time. With an rms down to 1~$\mu$Jy/beam at 1~GHz over 1,000 -- 5,000 deg$^2$ in 1 year (wide tier band 1/2)
and an rms down to 200~nJy/beam over 10 -- 30 deg$^2$ in 2000 hours (deep tier band 1/2), these surveys will directly detect faint radio-loud and radio-quiet AGN (down to a 1~GHz radio luminosity of about $2\times10^{23}$~W/Hz at $z=6$). For the first time, this will enable us to conduct detailed studies of the cosmic evolution of radio AGN activity to the cosmic dawn ($z\gtrsim6$), covering all environmental densities. 
}

\FullConference{
Advancing Astrophysics with the Square Kilometre Array\\
June 8-13, 2014\\
Giardini Naxos, Italy}

\newcommand{\skipthis}[1]{}

\def\lum               {$\mathrm{L}_\mathrm{1.4GHz}$}

\def\wh                {W/Hz}
\def\msol                {$\mathrm{M_\odot\,yr^{-1}}$}

\def\smo               {Smol\v{c}i\'{c}}
\def\f#1   {Fig.~\ref{#1}}
\def\comm#1   {{\tt (COMMENT: #1) }}

\begin{document}

\section{Introduction}

Galaxies are thought to evolve over time from an initial stage of blue star forming galaxies with spiral morphology towards quiescent red galaxies with spheroidal morphologies and the highest stellar masses \citep[e.g.,][]{faber07}. A galaxy evolves through interspersed episodes of intensive mass accretion onto the stellar body, as well as the central super-massive black hole (SMBH), creating a powerful active galactic nucleus \citep[AGN;][]{sanders96}. This is consistent with the $\Lambda$CDM paradigm, in which structure in the Universe grows hierarchically in such a way that small structures evolve into larger ones. In this context, faint (\lum < $10^{25}$~\wh ) but still radio-loud AGN  remain puzzling.\footnote{
For Type 1 AGN we here define radio loudness ($R'$) following \citet{white00} and \citet{ivezic02} as the logarithm of the ratio of radio-to-optical fluxes: 
$R'=\log\frac{F_\mathrm{radio}}{F_\mathrm{optical}} = 0.4(i-t)$ where $i$ is the optical $i$-band AB magnitude and $t=-2.5\log \left( \frac{F_\mathrm{1.4GHz}}{3631~Jy} \right)$ is the AB magnitude at 1.4~GHz. The adopted radio-loud vs.\ radio-quiet threshold is  $R'=1$. For Type 2 AGN we define radio-loud sources
as those with $q = \log{[(F_\mathrm{FIR}/3.75 \times 10^{12})/F_\mathrm{1.4GHz}]} < 1.7$, where $F_\mathrm{FIR}$ is 
the far-IR flux between $42.5~\mu$m and $122.5~\mu$m (e.g., Machalski \& Condon 1999;
Padovani et al. 2011) . 
}
 These faint radio loud AGN are found in red, quiescent galaxies that would not be identified as AGN at any other wavelength \citep[e.g.,][]{hickox09} and they do not seem to fit into the Unified Model for AGN \citep[e.g.,][]{hardcastle07}. They often reside at the bottom of the galaxy cluster/group potential wells and their radio-bright outflows heat the intra-cluster/group gas and the hot gas halo of the host galaxy \citep[e.g.][]{fabian12,best06}. This heating, deemed crucial in cosmological models of galaxy formation, has been termed `radio-mode' feedback \citep{granato04,croton06,bower06}. However, both on group/cluster and galaxy scales, feedback is still poorly understood.\footnote{We note that powerful (\lum > $10^{25}$~\wh ), high-excitation AGN, that fit into the Unified model for AGN, also can exert feedback through their jets \citep[e.g.][]{shabala11, zakamska14}. However, cosmological models assume that `radio-mode' feedback is cosmologically important through feedback exerted by galaxies that have reached their `quiescent' phase of black-hole accretion \citep[e.g.][]{croton06}. This `quiescent' galaxy phase is directly linked to low-excitation radio AGN \citep{smolcic09} predominantly found at low radio luminosities \citep[e.g.][]{kauffmann08, best12}.}

In the context of the most powerful AGN, Type 1 (broad line) AGN (quasars hereafter), are a population that experiences the most intense SMBH growth. Quasar winds associated with this intense SMBH growth are thought to quench their galactic star formation by expelling a fraction of the interstellar gas \citep[so called `quasar mode AGN feedback'; e.g.][]{hopkins06}. The existence of two, physically distinct, radio-loud (RL hereafter) and radio-quiet  (RQ hereafter)\footnote{
We here 
refer to radio luminous Type 1, broad line AGN as quasars.
} quasar populations is a long debated issue that has far-reaching implications for astrophysical models, including unified models for AGN and the evolution of star formation. Although the quasar radio-loudness distribution has been carefully studied in many different quasar samples over the past few decades \citep[e.g.][]{strittmatter80,ivezic02,white00,white07,cirasuolo03a,balokovic12}, there is still no definite understanding or consensus. The bimodality could imply two physically distinct types of quasars (pointing to e.g.\ different SMBH accretion/spin mechanisms or physically different sources of synchrotron emission) or be due to differing geometries \citep[e.g.][]{fanidakis11}. Furthermore, recent studies suggest that, in general,  radio emission in radio-quiet AGN may arise from star formation related processes \citep[e.g.][]{kimball11,padovani_etal11, condon12,bonzini14}. However, this is contradicted by \citet{white14}, who find evidence for black-hole accretion still making a significant contribution to the total radio emission. We note that such studies are subject to biases in terms of the luminosity and redshift ranges studied. For example, the deep-field work of \cite{bonzini14} is sensitive to fainter AGN which may be hosted in spiral galaxies which are likely to have ongoing star formation, whereas the wide-field work of \cite{kimball11} and \cite{condon12} detect the more luminous AGN, which are more likely to be hosted by massive ellipticals \citep[e.g.][]{Dunlop2003}.

In extragalactic radio surveys, two dominant galaxy populations are observed. These are AGN and star forming galaxies. At the bright end ($\gtrsim 1$ mJy), the GHz radio-bright sky consists mainly of ``classical'' radio
AGN, i.e.\ radio quasars and radio galaxies. Their radio emission is generated from the gravitational
energy associated with a SMBH and emitted through
relativistic jets of particles as synchrotron radiation. Below 1 mJy there is
an increasing contribution to the radio source population from massive star formation. In this case, synchrotron
emission is produced via relativistic plasma ejected from supernovae. However, such star forming galaxies (SFGs) appear not to be the only component of the faint radio sky, at least down
to $\sim 50~\mu$Jy at a few GHz \citep[e.g.,][]{gruppioni03,jarvisrawlings2004,simpson2006,smolcic08,mignano08,padovani09}. At the faint radio levels ($<1$~mJy) the source counts are still well populated by both RQ and RL AGN.

Sensitive radio continuum surveys, as will be provided with the SKA1 wide and deep tier band 1/2 surveys, are of extreme relevance for a variety of
reasons: (1)  Only deep radio observations trace AGN hosted by otherwise quiescent galaxies, thought to be the main drivers of the radio-mode feedback; 
(2) The least luminous RQ AGN reside typically in spiral
galaxies, which are still forming stars, and therefore are likely to provide
a vital contribution to our understanding of AGN -- galaxy co-evolution; (3) Radio observations are unaffected by absorption and therefore 
sensitive to all types of AGN, independently of obscuration and their orientation (i.e., Type
1s and Type 2s); (4) Finally and most importantly, sensitive
radio observations, that only the new-generation radio interferometers can provide, will start to detect the bulk ($\sim 90\%$) of the AGN population, currently missed by the majority of existing radio surveys \cite[e.g.][]{ivezic02,balokovic12}. 

For the remainder of the Chapter we assume the following SKA1 deep and wide tier band 1/2 survey characteristics. 	   
For the wide survey we assume an rms of $\sim1~\mu$Jy/beam over 1,000 -- 5,000~deg$^2$ reached in 1 year of observations, and for the deep survey we assume an rms of 0.2~$\mu$Jy/beam over 10 -- 30~deg$^2$ for 2,000~hours of observations at an observing frequency of 1~GHz (see \citealt{prandoni15}, for more details on the surveys).

\section{AGN activity in the faint radio sky}

\begin{figure}
\includegraphics[bb= 18 44 592 518, scale=0.7]{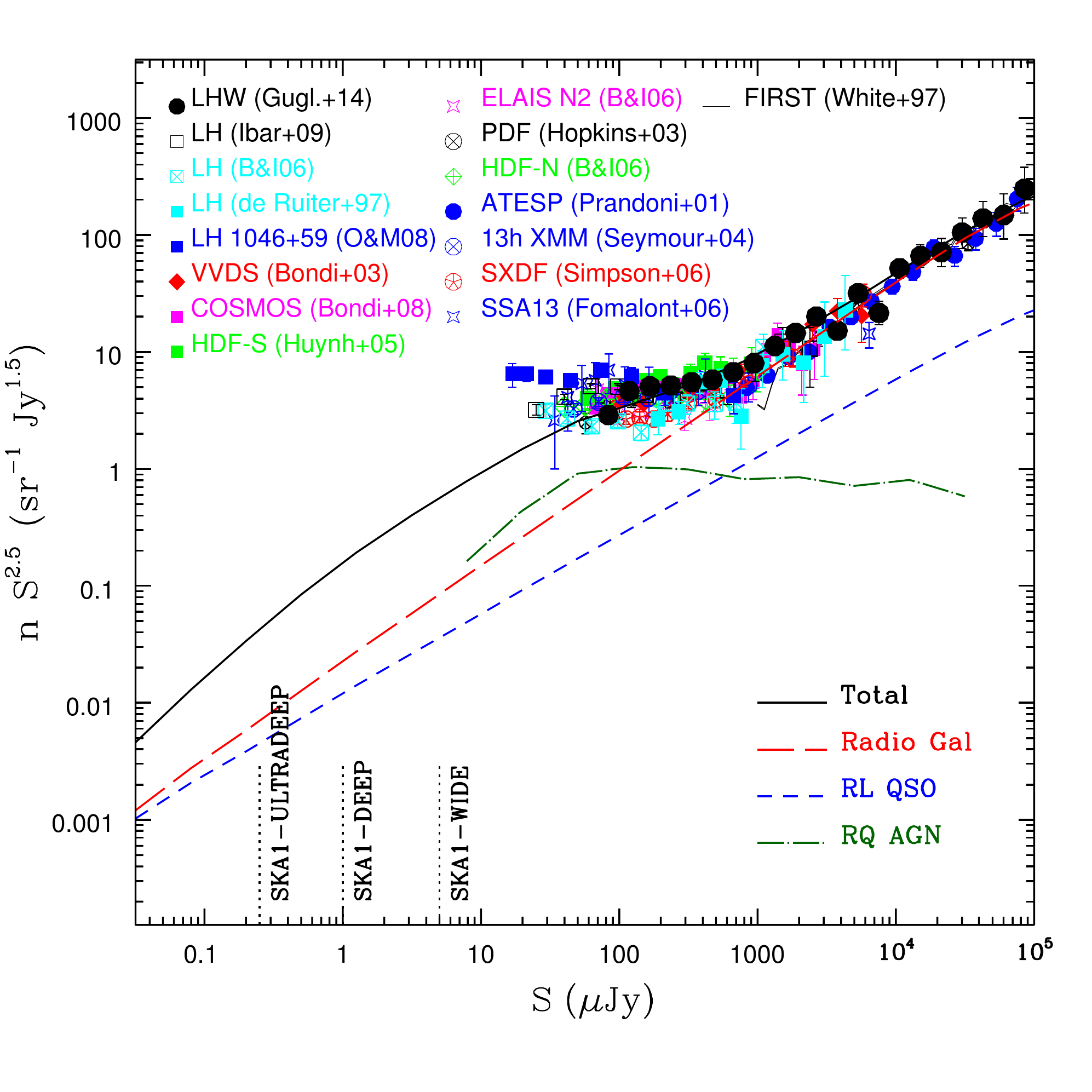}
\caption{A compilation of observed radio source counts and simulated counts, divided into radio-loud quasars, radio-quiet AGN, and radio galaxies (as indicated in the legend, the data are taken from Guglielmino et al., in preparation; \citealt{ibar09,biggs06,deruiter97,owen08,bondi03,bondi08,huynh05,biggs06,hopkins03,prandoni01,seymour04,simpson06,fomalont06,white97}). The simulated counts are taken from SKADS model \citep{wilman08}. Also indicated are the $5\sigma$ SKA1 wide, deep and Ultradeep limits.}
 \label{fig:draper}
\end{figure}

Radio source counts are the most straight-forward information drawn from a radio survey
and are commonly used to predict source counts in future deeper surveys. They flatten below 1 mJy and are generally
expected to decrease again at fainter fluxes \citep[e.g.][]{hopkins00,wilman08,condon12}.
In \f{fig:draper} \ we show a compilation of various survey results and the simulated results for presently unreached levels, separated for RL and RQ AGN.  Various methods, all relying on multi-wavelength data, have been employed  to separate the radio faint population into star forming and AGN galaxies \citep[e.g.][]{smolcic08, seymour08,simpson12,mcalpine13, bonzini13}. For example, using optical, IR, X-ray and radio data, \cite{bonzini13} disentangle the SFG, RQ, and RL AGN in the Extended {\it Chandra} deep Field South (ECDFS) survey \citep[$\sim 6~\mu$Jy rms noise in a 2.8" x 1.6" beam over 0.3
deg$^2$ containing 900 radio sources; see also][]{miller08,miller13,bonzini12,padovani_etal11}.

\begin{figure} \centering
\begin{minipage}{.5\textwidth}
  \centering
  \includegraphics[width=1.1\linewidth]{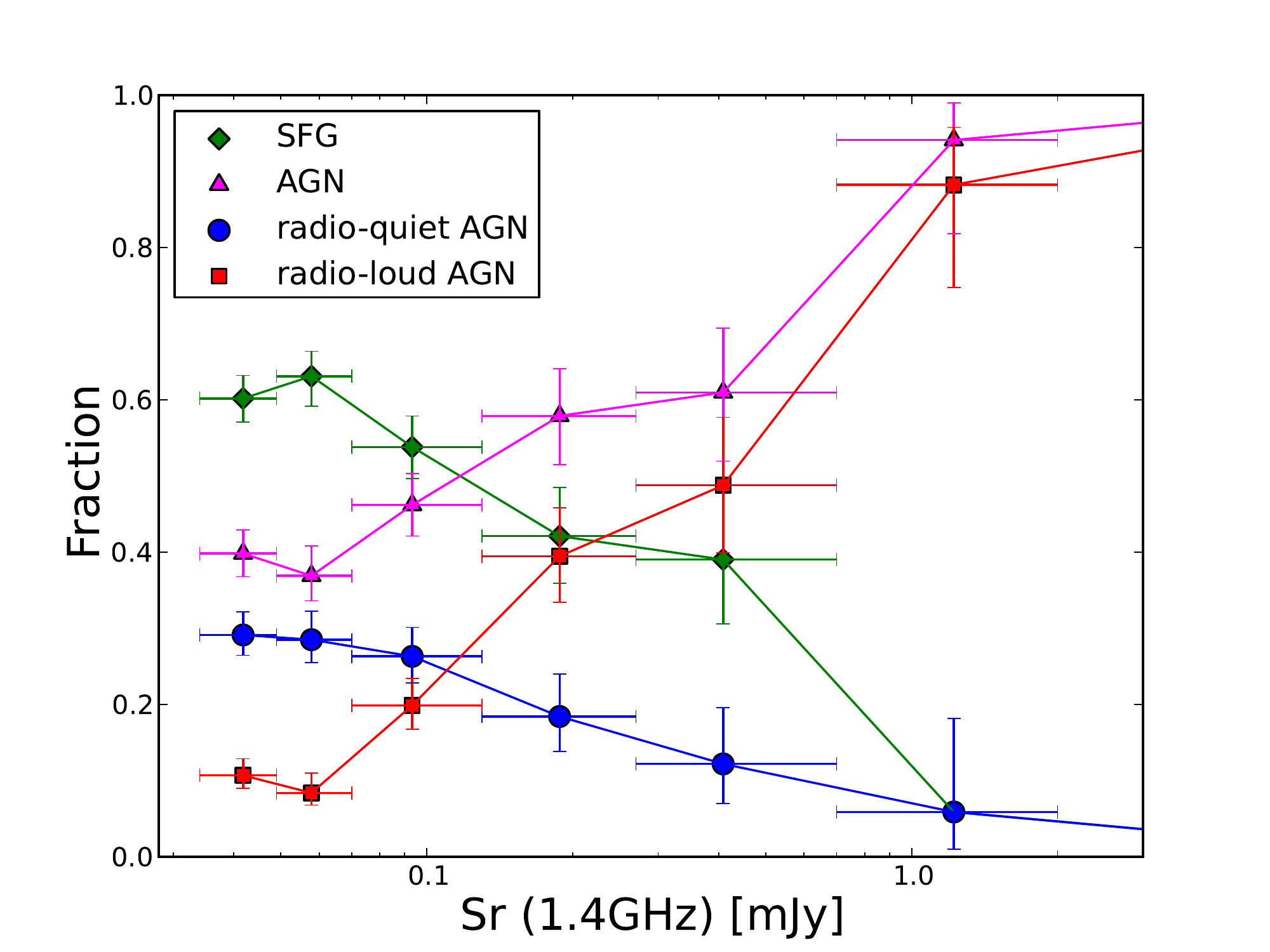}
\end{minipage}
\caption{
   Relative fraction of the various classes of E-CDFS sources as a function
 of radio flux density: SFG
   (green diamonds), all AGN (magenta triangles), radio-quiet AGN (blue
   circles) and radio-loud AGN (red squares). Adapted from
   \cite{bonzini13}.
    \label{fig:counts}
   }
\end{figure}

Fig. \ref{fig:counts}, adapted from \cite{bonzini13}, shows the relative fractions of 
the various radio source classes as a
function of radio flux density. As expected, AGN dominate at large flux
densities ($\gtrsim 1$ mJy). Above $\sim0.1$~mJy RL AGN are the predominant type of AGN, however
their fractional contribution steeply decreases towards lower flux densities. 
On the other hand, below 0.1 mJy the radio sky is dominated by star-formation-related processes, rather than RL and/or RQ AGN. Thus, deep observations of large areas of the sky are needed to assemble statistically-sound samples of the faintest RQ and RL AGN and to study their physical properties and cosmic evolution. Such surveys will be provided by  SKA1 and SKA2.
At the above assumed depth and area of the wide (5000 deg$^2$, rms~$\sim1~\mu$Jy/beam) and deep (30 deg$^2$, rms~$\sim0.2~\mu$Jy/beam) tier SKA1 surveys, about $3\times10^7$ and $4\times10^5$ AGN, respectively, are predicted  
using the SKADS \citep{wilman08,wilman10} simulations of the extragalactic radio sky based on models of the evolution of the radio luminosity function. This is orders of magnitude larger than the number of sources detected in the deepest radio surveys to-date, e.g. COSMOS \citep[rms~$\sim10-15~\mu$Jy/beam, 2 deg$^2$, $\sim2,500$ sources,][]{schinnerer07,schinnerer10} and ECDFS \citep[rms~$\sim6~\mu$Jy/beam, 0.3 deg$^2$, $\sim900$ sources,][]{miller08,miller13}. For comparison,  MeerKAT-MIGHTEE
(rms~$\sim1~\mu$Jy/beam, 35 deg$^2$, 1.4~GHz) is expected to detect about 174,000 AGN, while ASKAP-EMU (rms~$\sim10~\mu$Jy/beam, 30,000 deg$^2$, 1.4~GHz) will detect $\sim 2.8\times10^7$ AGN.

\subsection{Radio emission in radio-quiet AGN}

RQ AGN are characterized
by relatively low radio-to-optical flux density ratios and radio powers and, until recently, have been found predominantly in optically selected samples.
We note that the distinction between RQ and RL AGN is
not simply a matter of semantics - the two classes represent intrinsically
different objects. RL AGN emit most of their energy over the entire 
electromagnetic spectrum non-thermally
and in association with powerful relativistic jets. The multi-wavelength 
emission of RQ AGN is dominated by thermal emission and is 
related to the accretion disk. 
The exact mechanism responsible for radio emission in RQ AGN has been a matter of
debate for the past fifty years. Explanations have included, for example, a scaled down
version of the RL AGN mechanism \citep[e.g.,][]{miller93,ulvestad05},
and star formation \citep{sopp91}.

There is still no clear consensus on the existence of a bimodality in the radio-loudness distribution of quasars. 
A bimodality would imply two physically distinct types of quasars in the Universe and would point 
to, for instance, different SMBH accretion/spin mechanisms, geometries or physical properties \citep[e.g.][]{fanidakis11,kimball11,condon13}. 
This  issue is still open mainly due
to the overwhelmingly high fraction of RQ quasars that
regularly go undetected in radio surveys. As shown in \f{fig:qso} \
all current radio surveys (even the deepest ones) barely sample even the loudest end of the radio-quiet part of the distribution. Hence,
as only $\sim10\%$ of optically selected quasars are RL
\citep[e.g.][]{ivezic02}, this means that a major fraction of quasars still remains
  undetected and unexplored at radio wavelengths \citep[see also][]{balokovic12}. Only observations of large sky areas to the depths reachable with the SKA1 deep and wide surveys will directly reveal the radio properties of roughly 90\% of optically detected quasars.

Recent studies of radio-quiet AGN suggest different radio emitting mechanisms in RL and RQ AGN \citep[e.g.][]{kimball11,padovani_etal11,condon13,bonzini13}.
For example, investigating the cosmic evolution of RQ AGN and SFGs, the results presented in \cite{padovani_etal11} 
suggest very close ties between star formation and radio emission in RQ AGN at $z \sim 1.5 - 2$. They find that the evolution of RQ AGN is similar to that of SFG \citep[see also][and references therein]{smolcic_etal09a} and that their luminosity function appears to be an extension
of the SFG LF \citep[see also][]{kimball11,condon13}. 
If RQ AGN were simply scaled-down versions of RL AGN, it could be expected that they share the evolutionary properties of the latter and their
luminosity function should also be an  extrapolation of the RL luminosity function at low
powers, however this does  not appear to be the case (e.g. \citealt{fernandes11}).
This has prompted studies of the emission mechanism of synchrotron radiation from RQ AGN. For example, \cite{kimball11,padovani_etal11,condon13,bonzini13} suggest that star formation in the host galaxy of RQ AGN may be the dominant contributor to the radio continuum emission, whereas \cite{white14} compare the radio emission from a sample of faint radio-quiet quasars with the massive galaxy population, and suggest that the radio emission in optically-selected radio-quiet quasars is consistent with being due to the AGN. These studies envelope different luminosity (both optical and radio) and redshift ranges, and some of the differences may be attributable to to the diversity of the samples.
Furthermore, a close link between star formation and radio emission in RQ AGN is further affirmed by the 
comparison of the star formation rates (SFRs) derived from the far-IR luminosities and the radio luminosities, assuming that all the radio emission is due to star formation \citep{bonzini14}. 
For RQ AGN and SFGs, the two SFR estimates are  consistent. For RL AGN, the agreement is poor due to the large contribution of the
relativistic jet to their radio luminosity \citep{moric10,bonzini14}. Another intriguing possibility is that both AGN and SF processes contribute to the total radio (and IR) emission, in some relative proportion \citep[e.g.][]{moric10}. Seyfert 2 galaxies are a well-established example in the local Universe \citep[e.g.][]{roy98} and recent studies indicate that composite AGN/SF systems may constitute a significant fraction of the galaxy population at high redshifts \citep[e.g.][]{daddi07,gruppioni11,delmoro13}. 

The SKA1 wide and deep surveys, in conjunction with multi-wavelength data will provide the basis to resolve the long-standing quasar radio-loudness dichotomy,  
and the possible interplay between coexisting AGN and star formation phenomena. 
 However, only once star-formation and AGN activity are reliably separated,
and/or their fractional contribution to individual sources is determined, it is possible to derive unbiased radio
luminosity functions and disentangle the contributions of each type of activity over cosmic time 
via direct radio detections and stacking in the radio map \citep[e.g.][]{smolcic08,karim11,zwart14}.

\begin{figure}
\begin{center}
\includegraphics[bb=  104 360 486 812, scale=0.7]{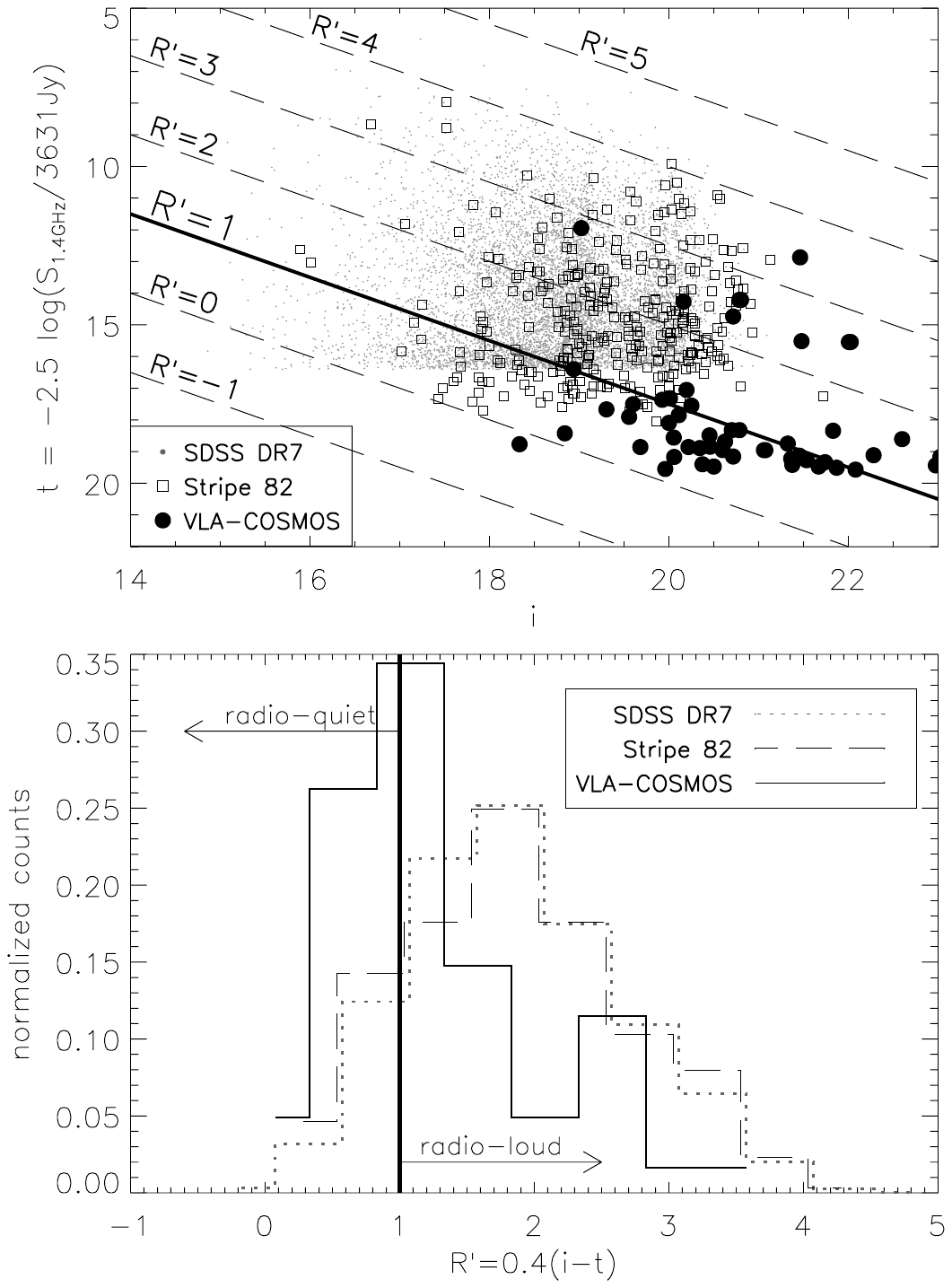}
\caption{Top: $i$-band vs.\ 1.4~GHz
        radio magnitude ($t$) distribution for optically selected
        quasars (i.e.\ broad line AGN) drawn from current
        state-of-the-art surveys: SDSS DR7 - FIRST \citep[$\sim9,000$ deg$^2$, $\mathrm{S_{1.4GHz}}\gtrsim1$~mJy;][]{schneider10},
        Stripe 82 \citep[92 deg$^2$, $\mathrm{S_{1.4GHz}}\gtrsim260~\mu$Jy;][]
        {hodge11}, and COSMOS \citep[2 deg$^2$,
        $\mathrm{S_{1.4GHz}}\gtrsim12~\mu$Jy;][]{schinnerer07,
        schinnerer10,lilly07,lilly09}. The bottom panel shows the
        distribution of radio loudness, $R'=0.4(i-t)$, for quasars in
        these three surveys. 
 \label{fig:qso}
  }
  \end{center}
\end{figure}

\section{Radio-loud AGN: Relevance for feedback in massive galaxy formation}

By now, negative AGN feedback has become a standard
ingredient in semi-analytic models and is required to reproduce the observed
galaxy properties \citep[e.g.][]{granato04,croton06,bower06,sijacki07}.
 In the
models, this type of feedback, referred to as radio-mode feedback, is related to radio AGN outflows  as the main source that heats the gas halo surrounding a
massive galaxy. This heating thereby quenches the star formation and limits
growth, thus avoiding the creation of overly high-mass galaxies.  
A detailed description of radio feedback is given in \citet{mcalpine15}.

The first observational support for AGN feedback was found using the combination of radio and X-ray data by
\cite{mcnamara00}, and \cite{best06} quantitatively showed that, in the local
Universe, radio outflows may indeed balance the radiative cooling of
the hot gas surrounding elliptical galaxies.  Furthermore, it has been
both theoretically postulated and observationally supported that this
`radio-mode' heating occurs during a quiescent phase of
  black-hole accretion (presumably via advection dominated accretion flows) and manifests as low-power radio AGN activity \citep[\lum~$<10^{25}$~\wh ;][]{evans06,hardcastle06,hardcastle07,kauffmann08,smolcic_etal09b,smolcic09,smolcic11}. Such low-power radio AGN, and the evolution of their comoving volume density (i.e.\ their radio luminosity function) through cosmic
times, can be studied in detail only via simultaneously deep and large radio surveys with supplementary
panchromatic data \citep{sadler07,donoso09,smolcic_etal09a,simpson12,mcalpine13}.

\begin{figure}
\includegraphics[bb=54 390 486 792]{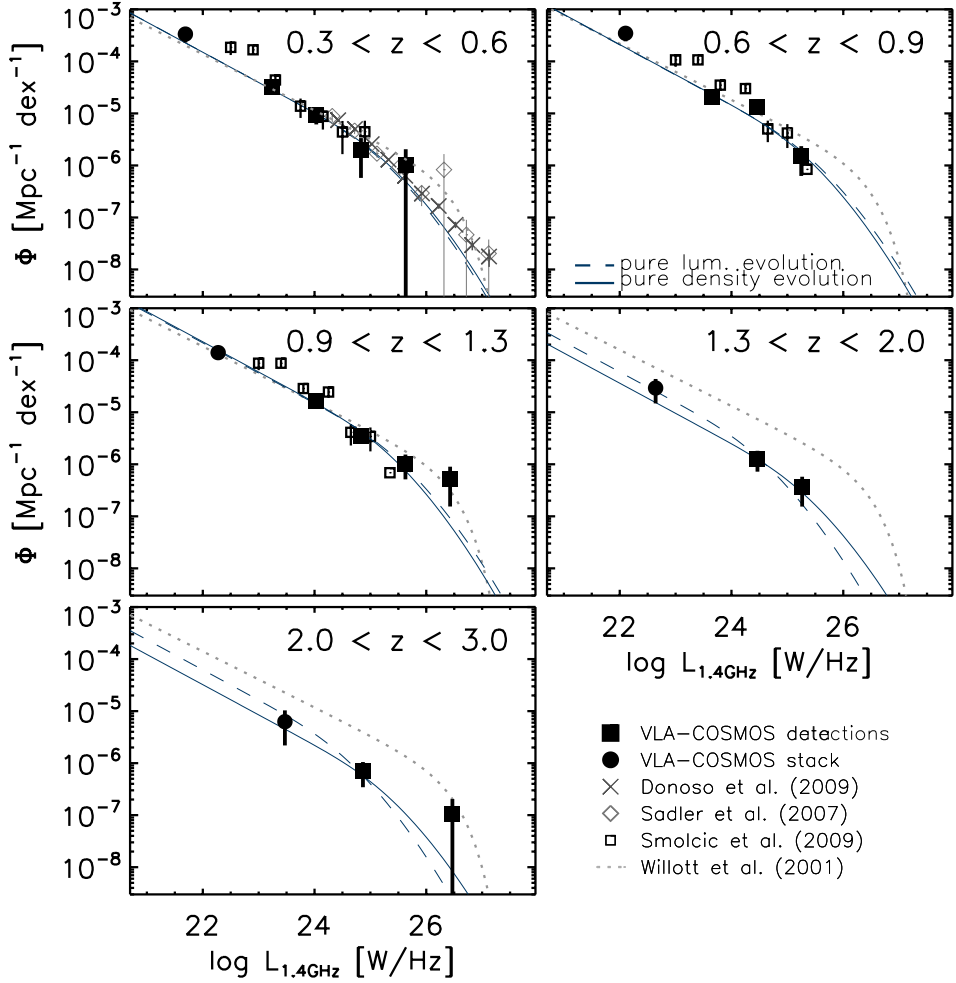}
\caption{Radio luminosity functions (LFs) in five redshift bins out to $z=3$ for quiescent galaxies in the
  COSMOS field with stellar
  masses $\mathrm{M_*}\gtrsim3\times10^{10}$~\msol \ \citep[][filled circles and squares]{ilbert10}. The volume
  densities derived based on a
  volume limited radio detected sample are shown by the filled black squares,
  while those based on stacked data
  are shown by the filled black circles. The (full and dashed) lines show the best fit
  evolution to the COSMOS data in a given redshift range (blue
  curves)
  using the \cite{sadler02} local LF (dashed line: pure luminosity evolution;
  full line: pure density
  evolution). Various results from the literature, indicated in the legend, are also shown. \label{fig:lfs}
  }
\end{figure}

\begin{figure}
\begin{center}
\includegraphics[bb=70 615 406 792, scale=0.7]{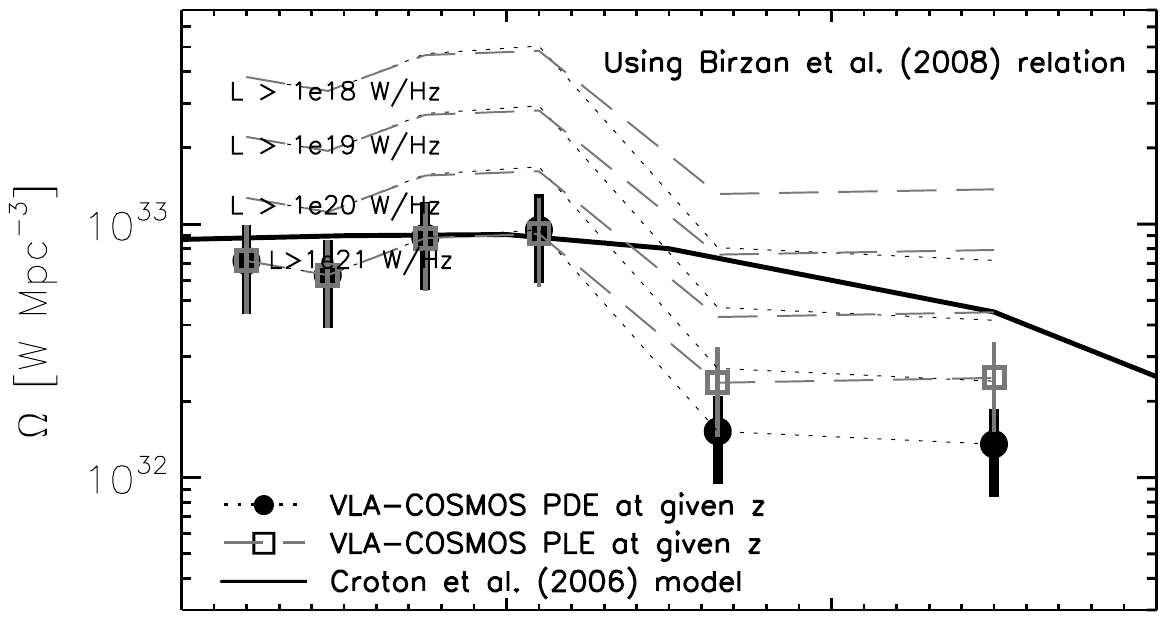}\\
\includegraphics[bb=70 570 406 792, scale=0.7]{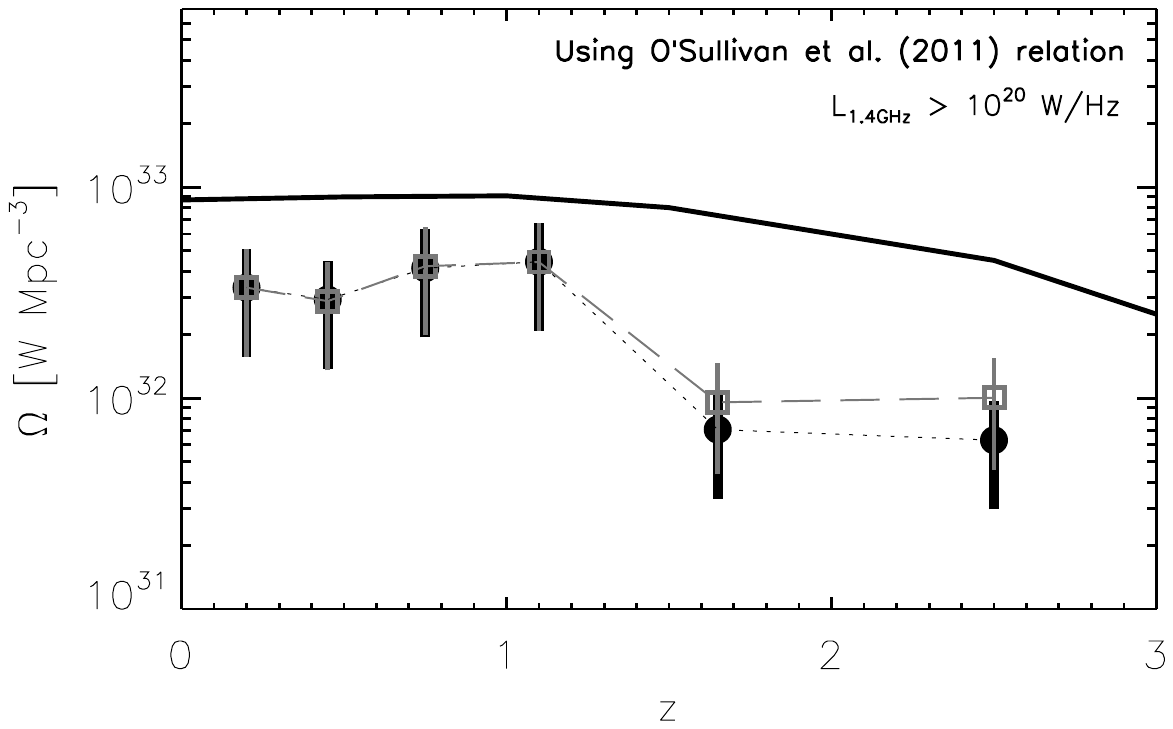}
\caption{ Comoving volume averaged heating rate ($\Omega$) performed by
  quiescent massive galaxies in the COSMOS field as a function of redshift. $\Omega$ was
  derived by integrating the kinetic power per comoving volume over
  radio luminosity \citep[see][for details]{smolcic_etal09a}. The COSMOS data points were derived by assuming pure density
  (filled circles), and pure luminosity (open squares) evolution best
  fit to the data in a given redshift bin. The upper and lower panels show $\Omega$ when using the relation between kinetic and 1.4~GHz radio luminosity from \cite{birzan08} and \cite{osullivan11}, respectively. The (dotted and dashed)
  lines in the upper panel illustrate the discrepancy in   $\Omega$ if various lower limit integral values are assumed
  (indicated in the panel). No such discrepancy is present when using the \cite{osullivan11} relation. The thick solid line in both panels shows the `radio-mode feedback' heating rate drawn from the \cite{croton06} cosmological model, and
  required to reproduce observed galaxy properties. 
  \label{fig:heat}}
\end{center}
\end{figure}

\begin{figure}
\includegraphics[bb=60 0 566 425, scale=0.27]{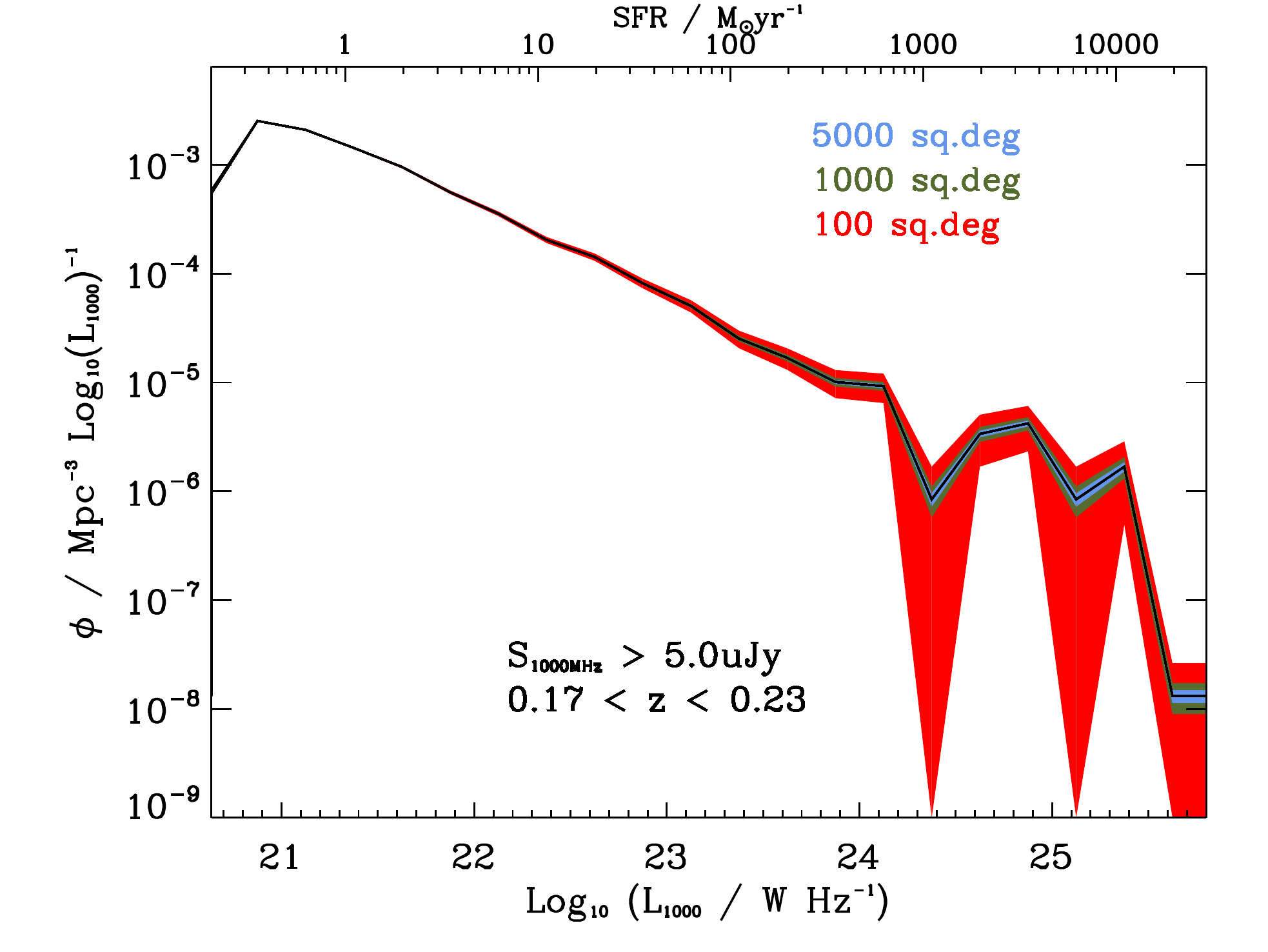}
\includegraphics[bb=40 0 566 425, scale=0.27]{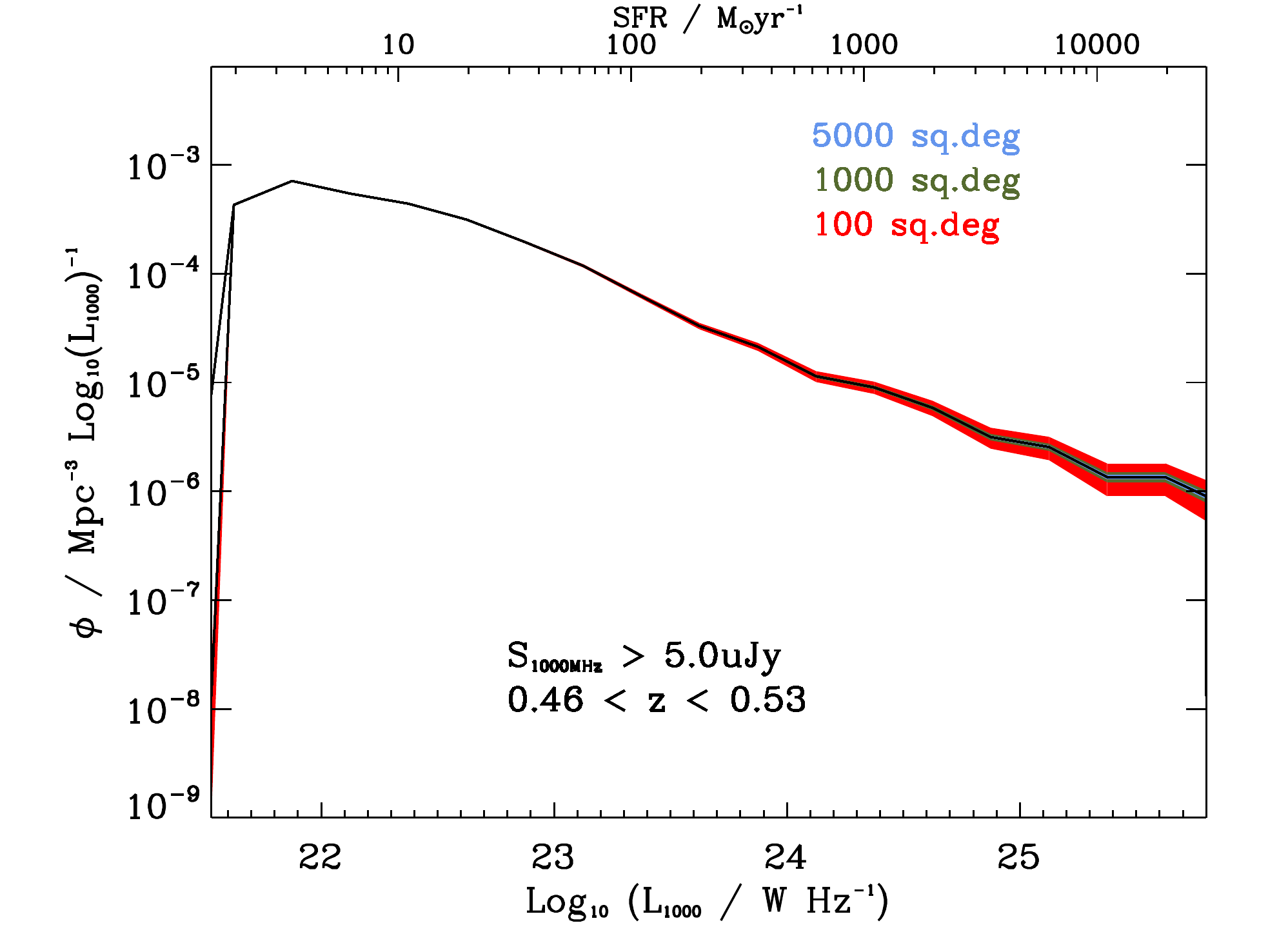}
\includegraphics[bb=40 0 566 425, scale=0.27]{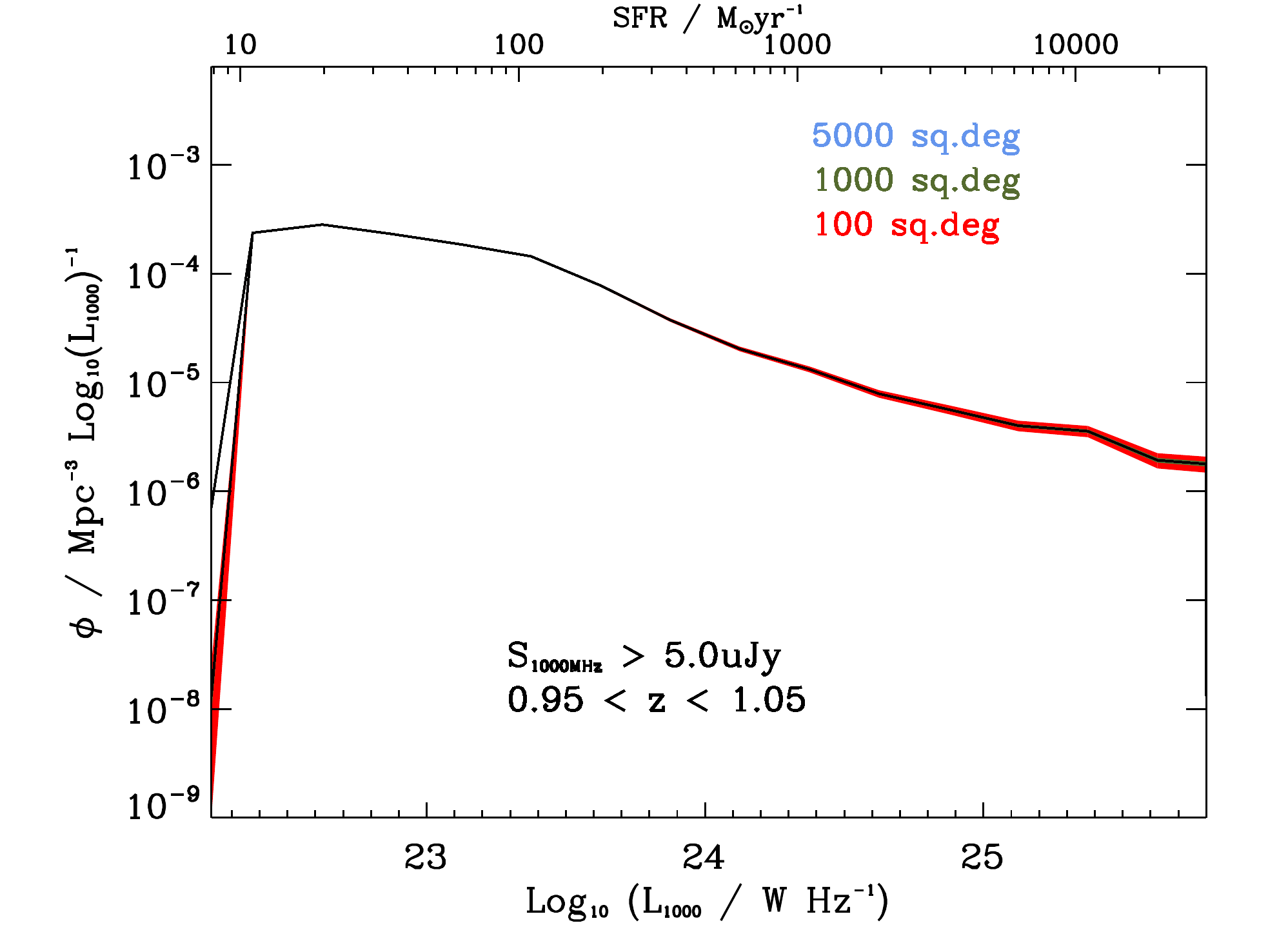}\\
\includegraphics[bb=60 0 566 425, scale=0.27]{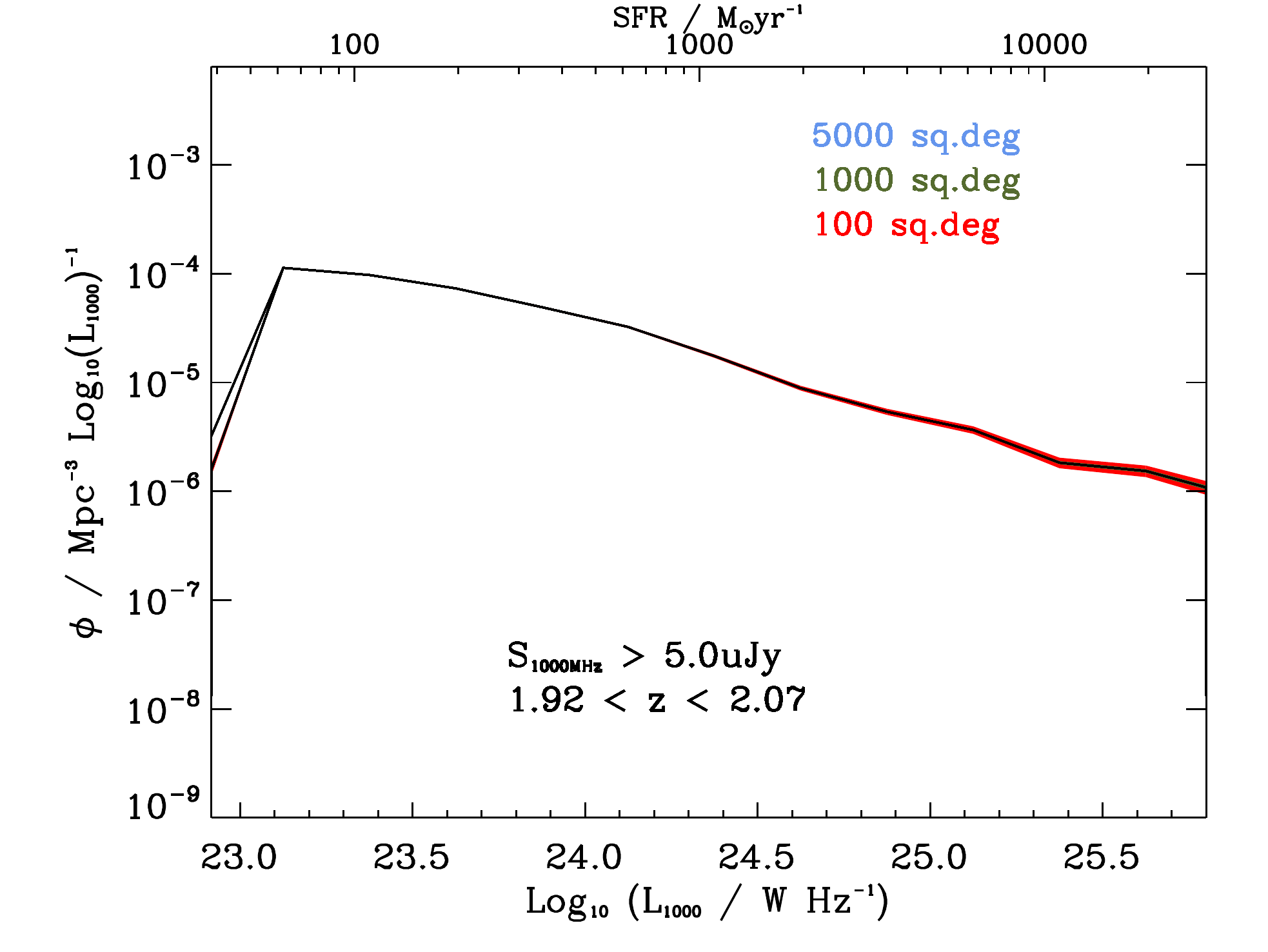}
\includegraphics[bb=40 0 566 425, scale=0.27]{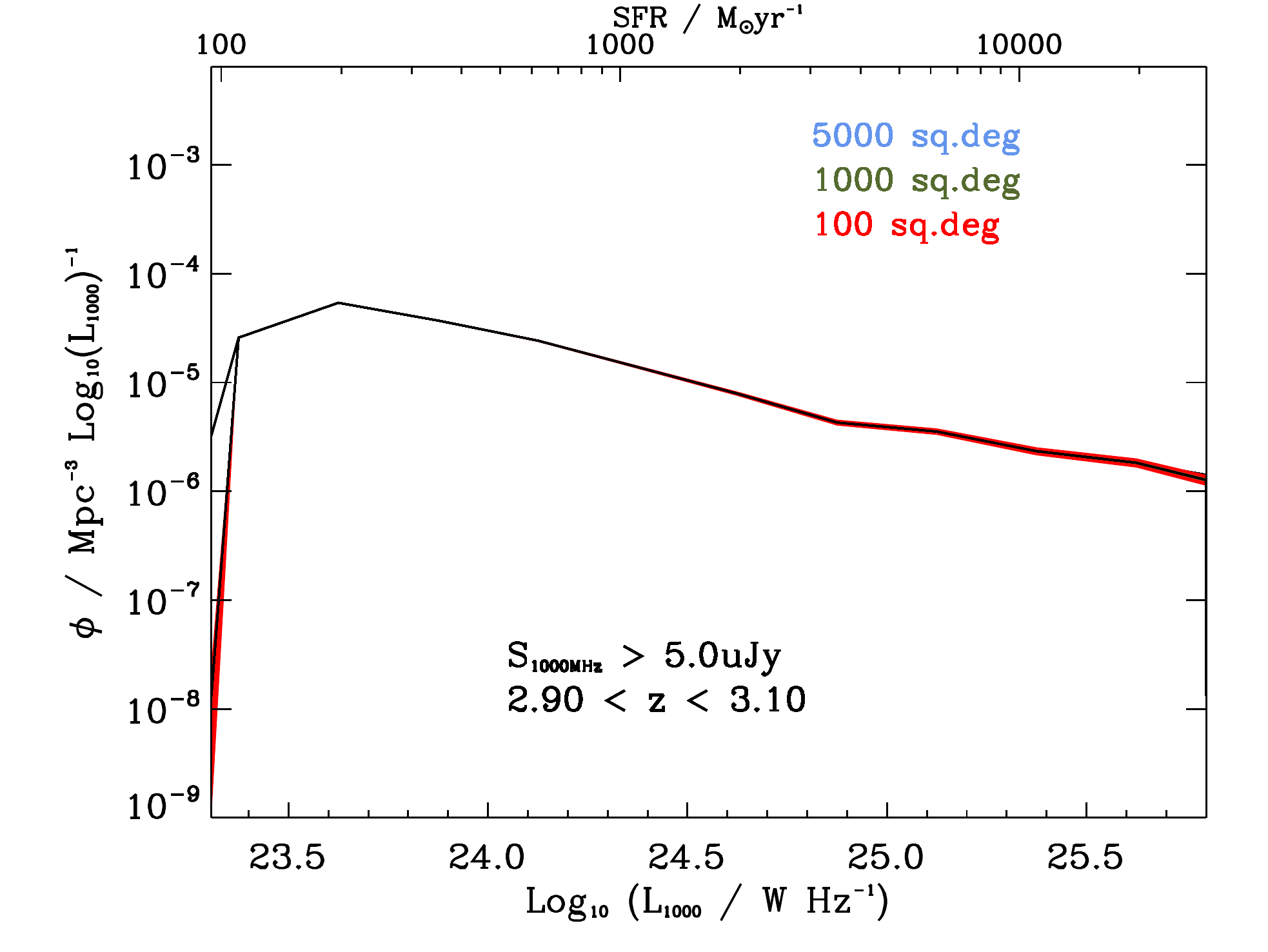}
\includegraphics[bb=40 0 566 425, scale=0.27]{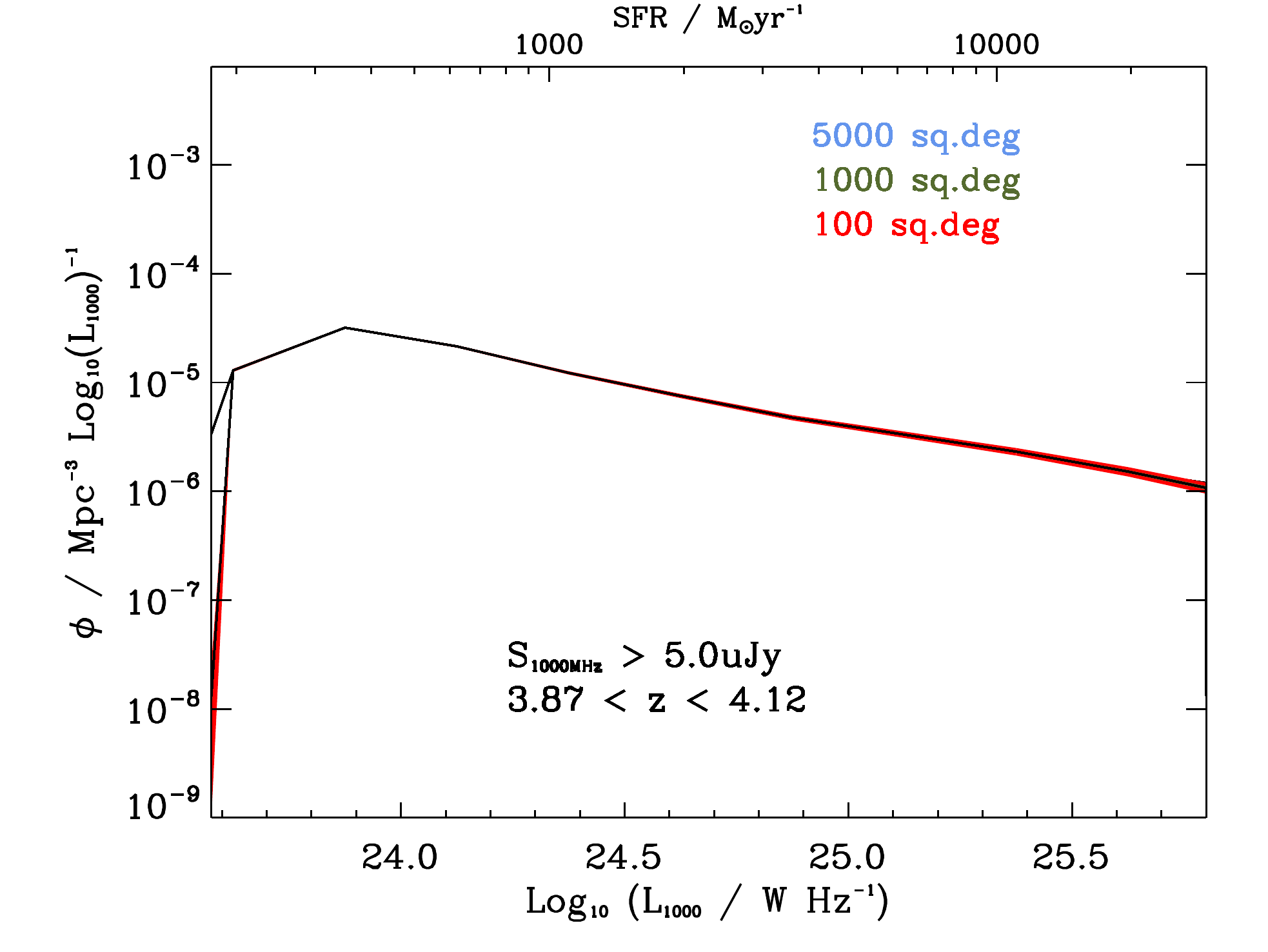}\\
\includegraphics[bb=60 0 566 425, scale=0.27]{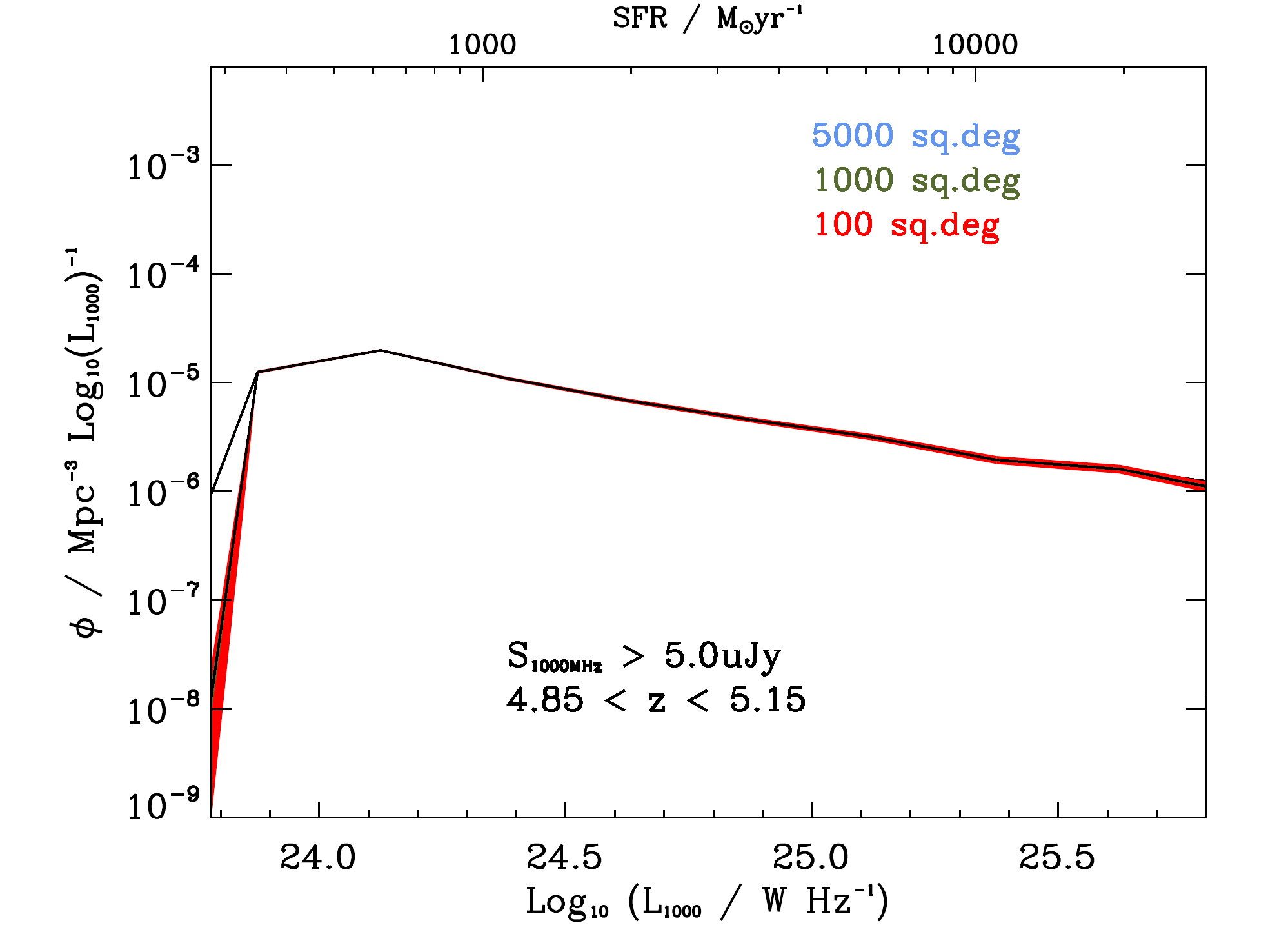}
\includegraphics[bb=40 0 566 425, scale=0.27]{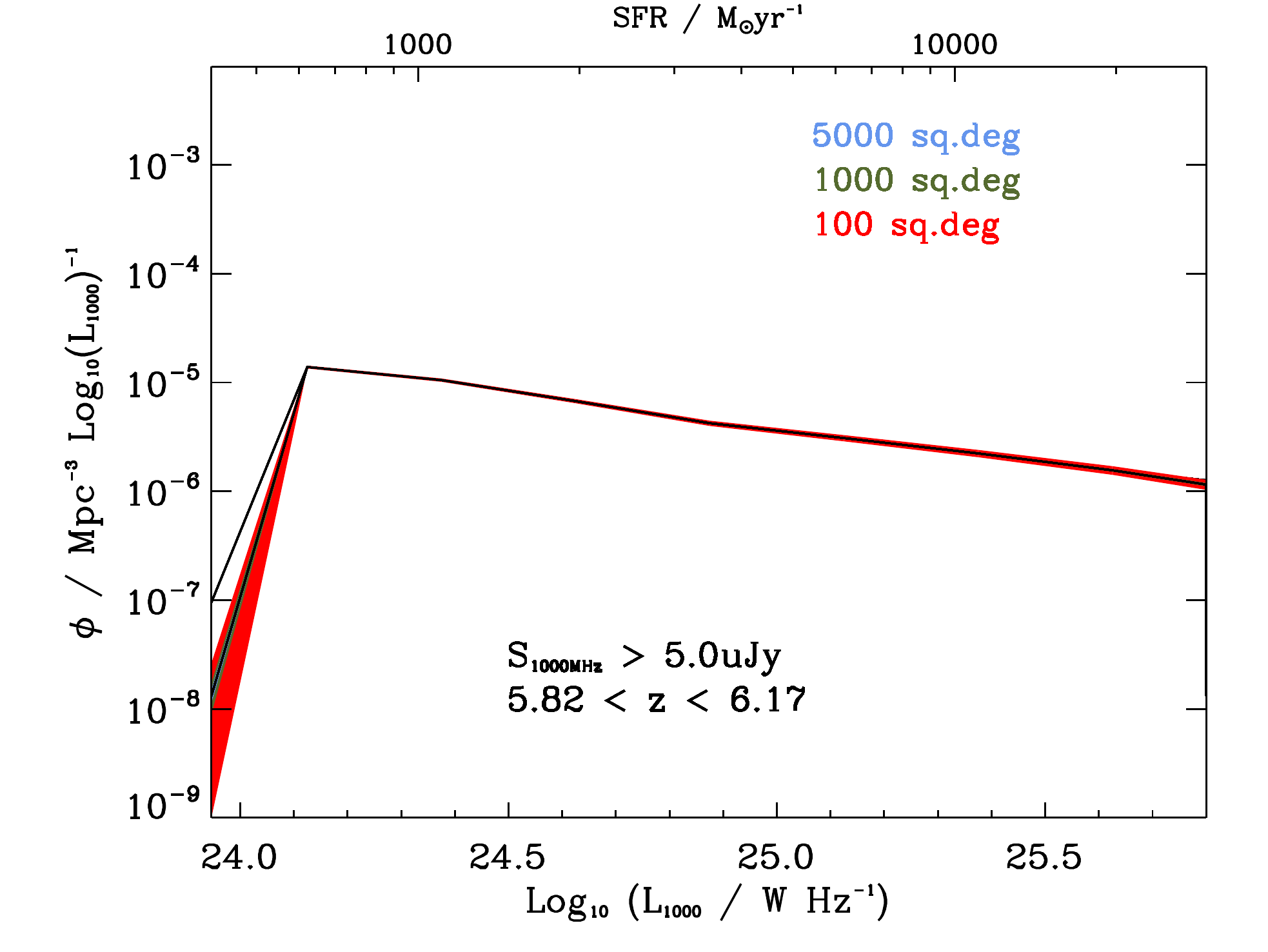}
\includegraphics[bb=40 0 566 425, scale=0.27]{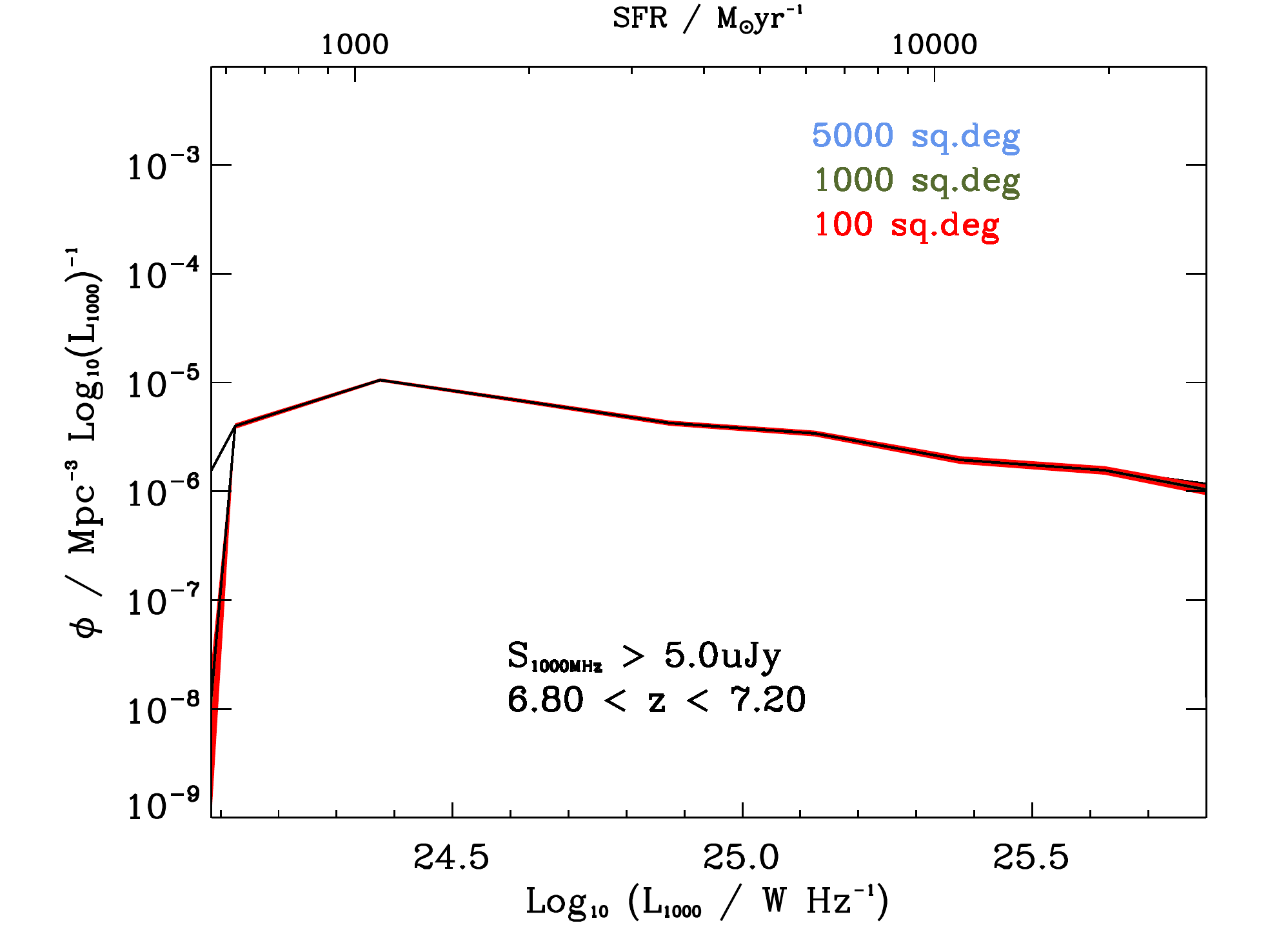}
\caption{Radio AGN luminosity functions in various redshift bins (indicated in the panels) predicted for the wide SKA1 tier based on the SKADS simulations \citep{wilman08,wilman10}. 
 \label{fig:lfsSKA1}
  }
\end{figure}

\begin{figure}
\includegraphics[bb=60 0 566 425, scale=0.27]{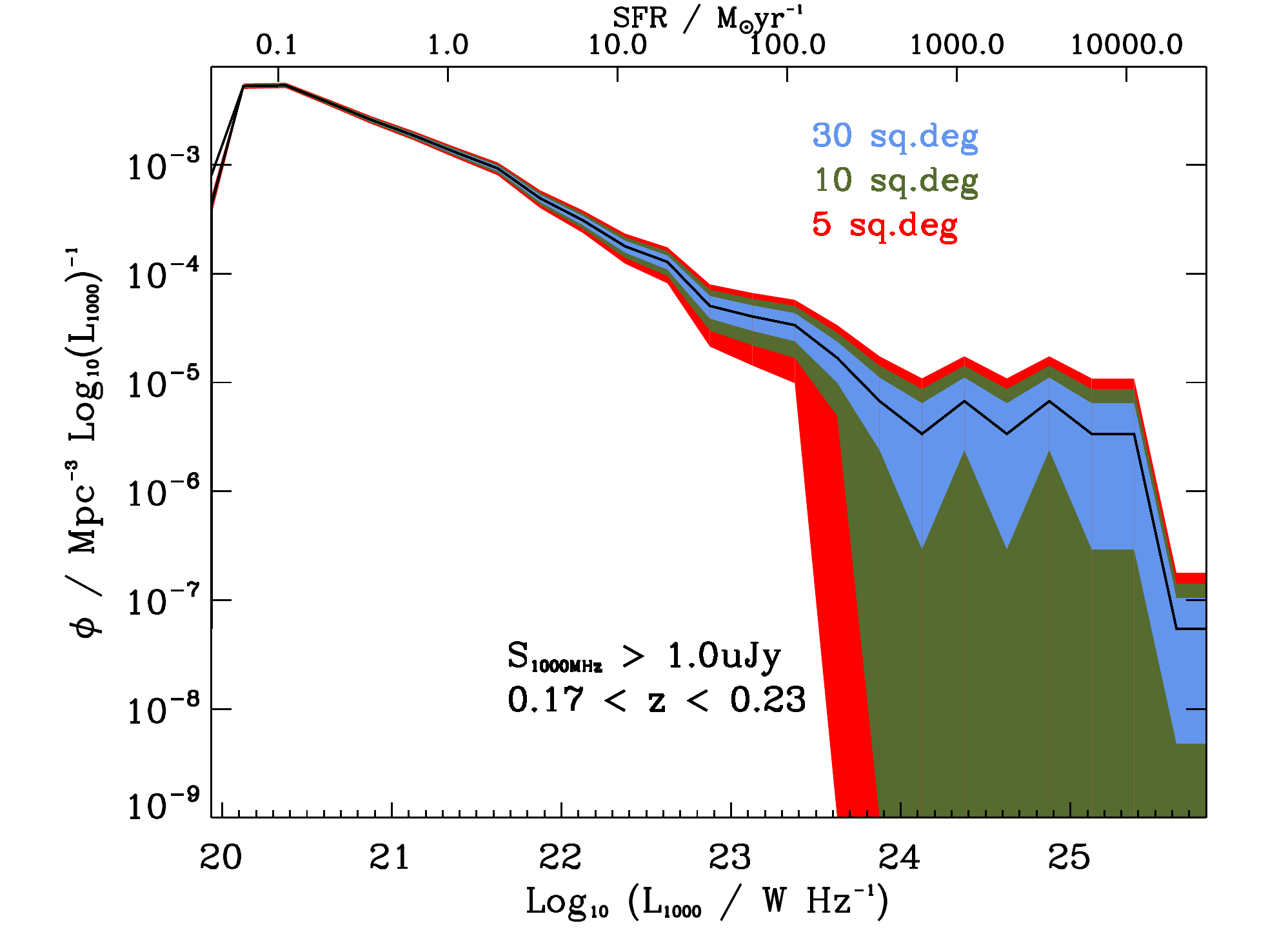}
\includegraphics[bb=40 0 566 425, scale=0.27]{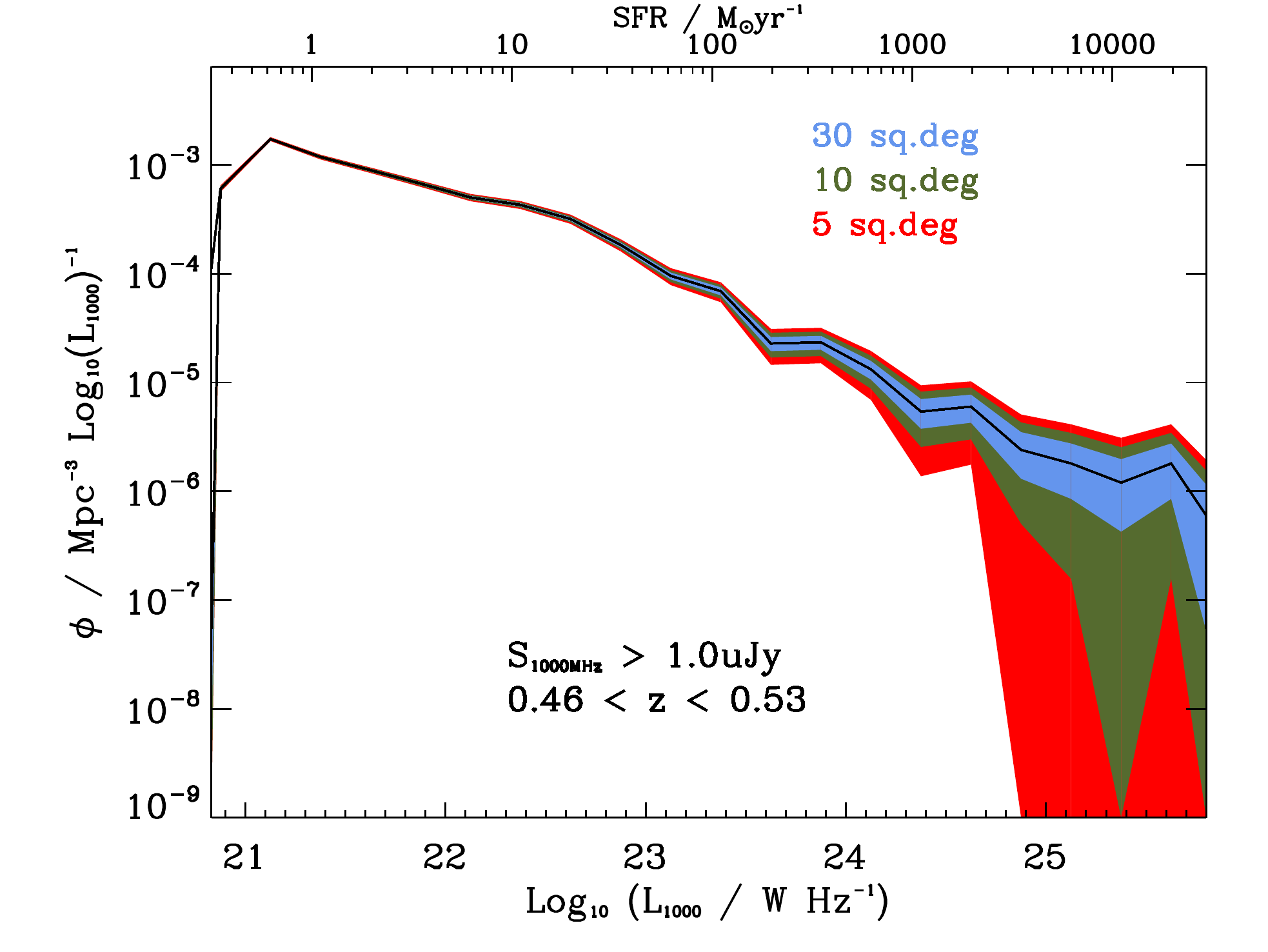}
\includegraphics[bb=40 0 566 425, scale=0.27]{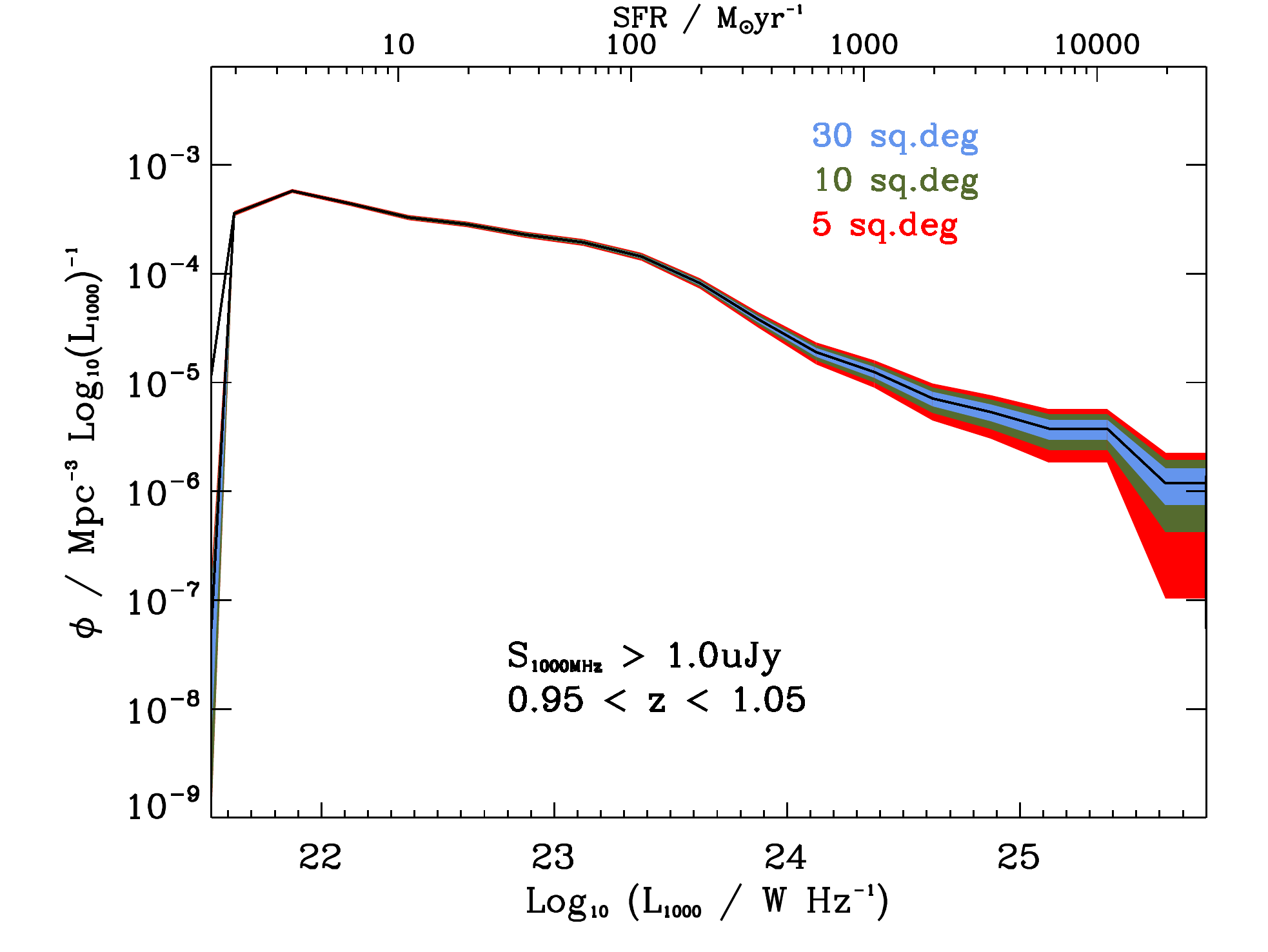}\\
\includegraphics[bb=60 0 566 425, scale=0.27]{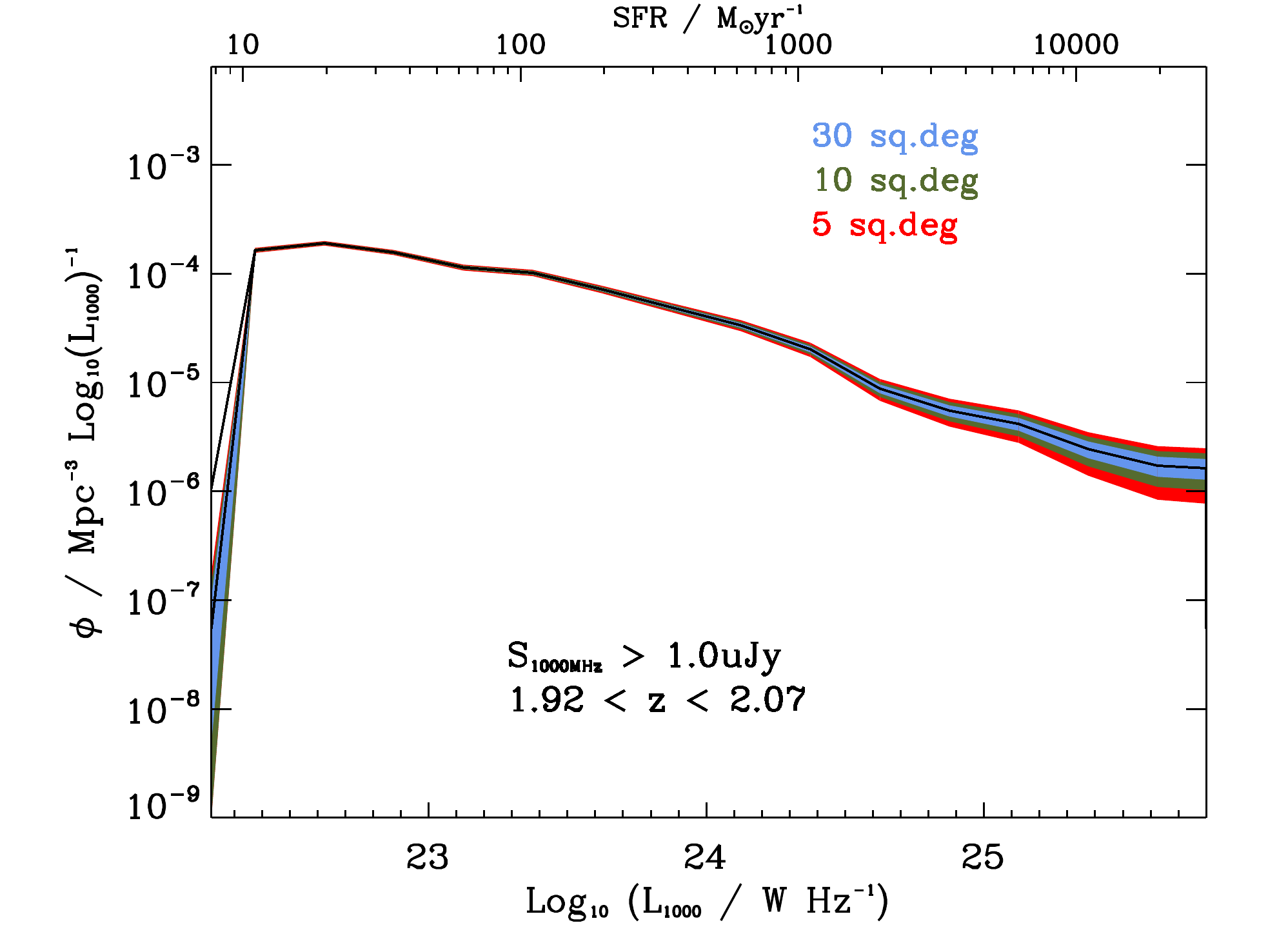}
\includegraphics[bb=40 0 566 425, scale=0.27]{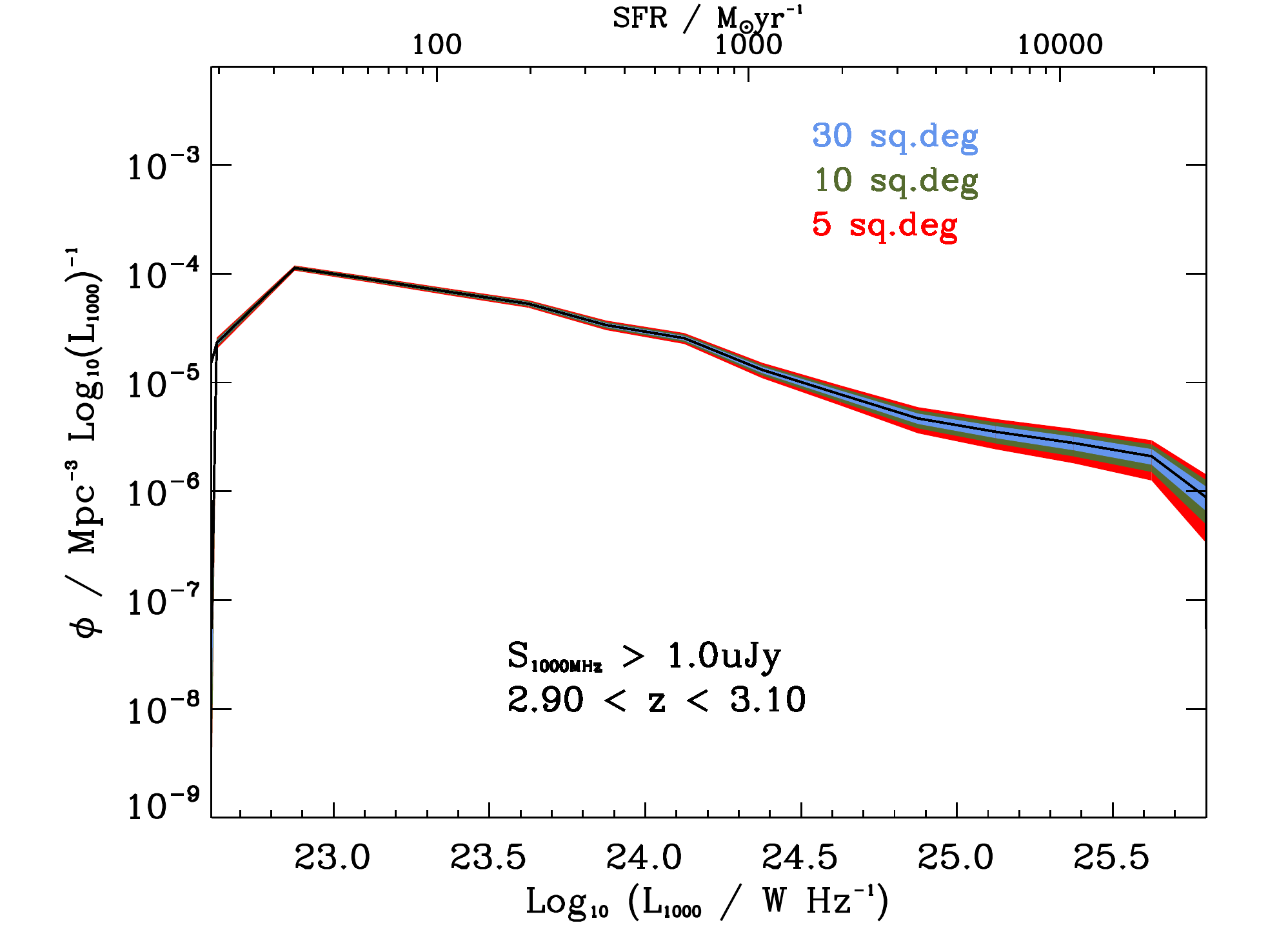}
\includegraphics[bb=40 0 566 425, scale=0.27]{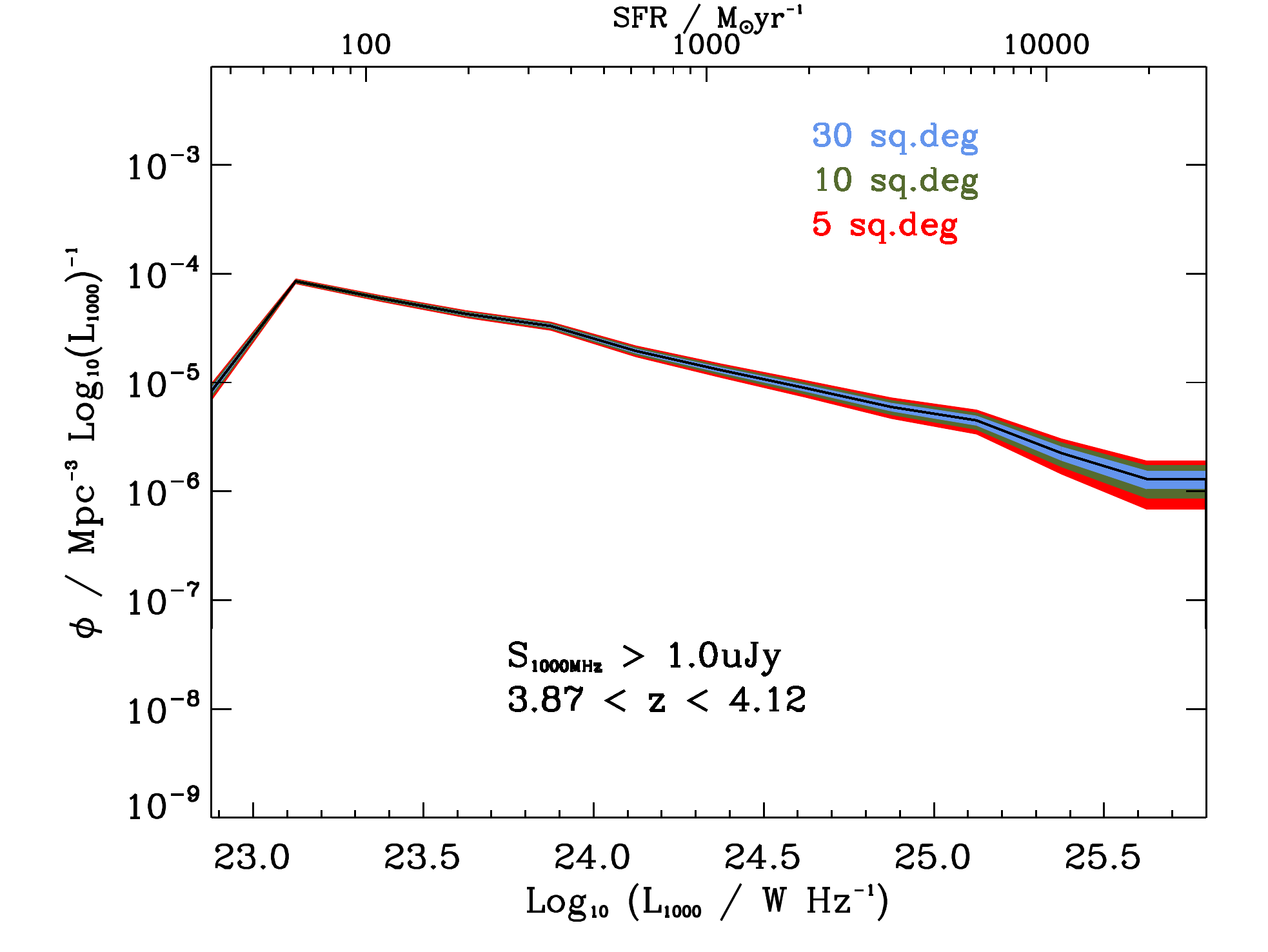}\\
\includegraphics[bb=60 0 566 425, scale=0.27]{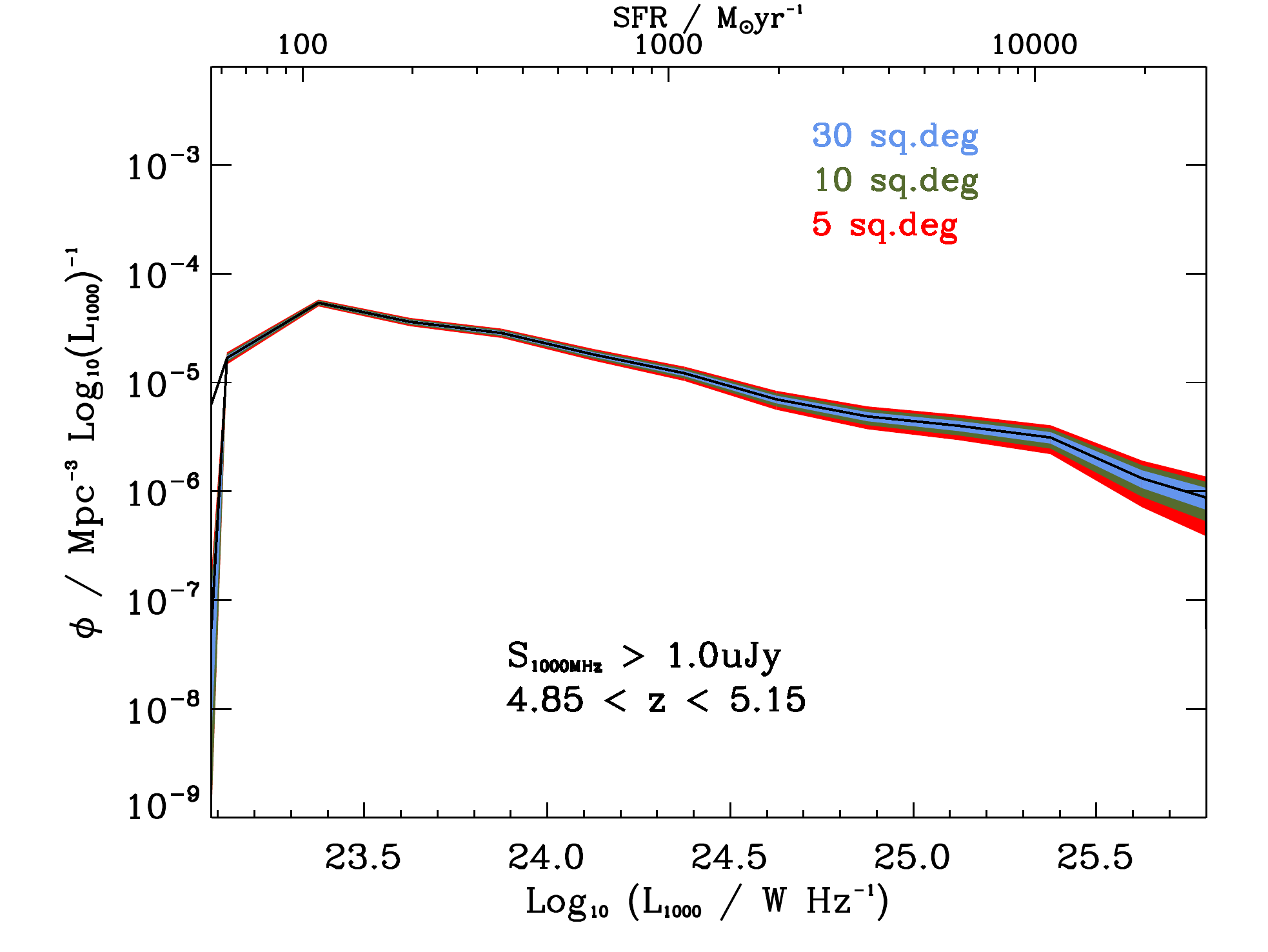}
\includegraphics[bb=40 0 566 425, scale=0.27]{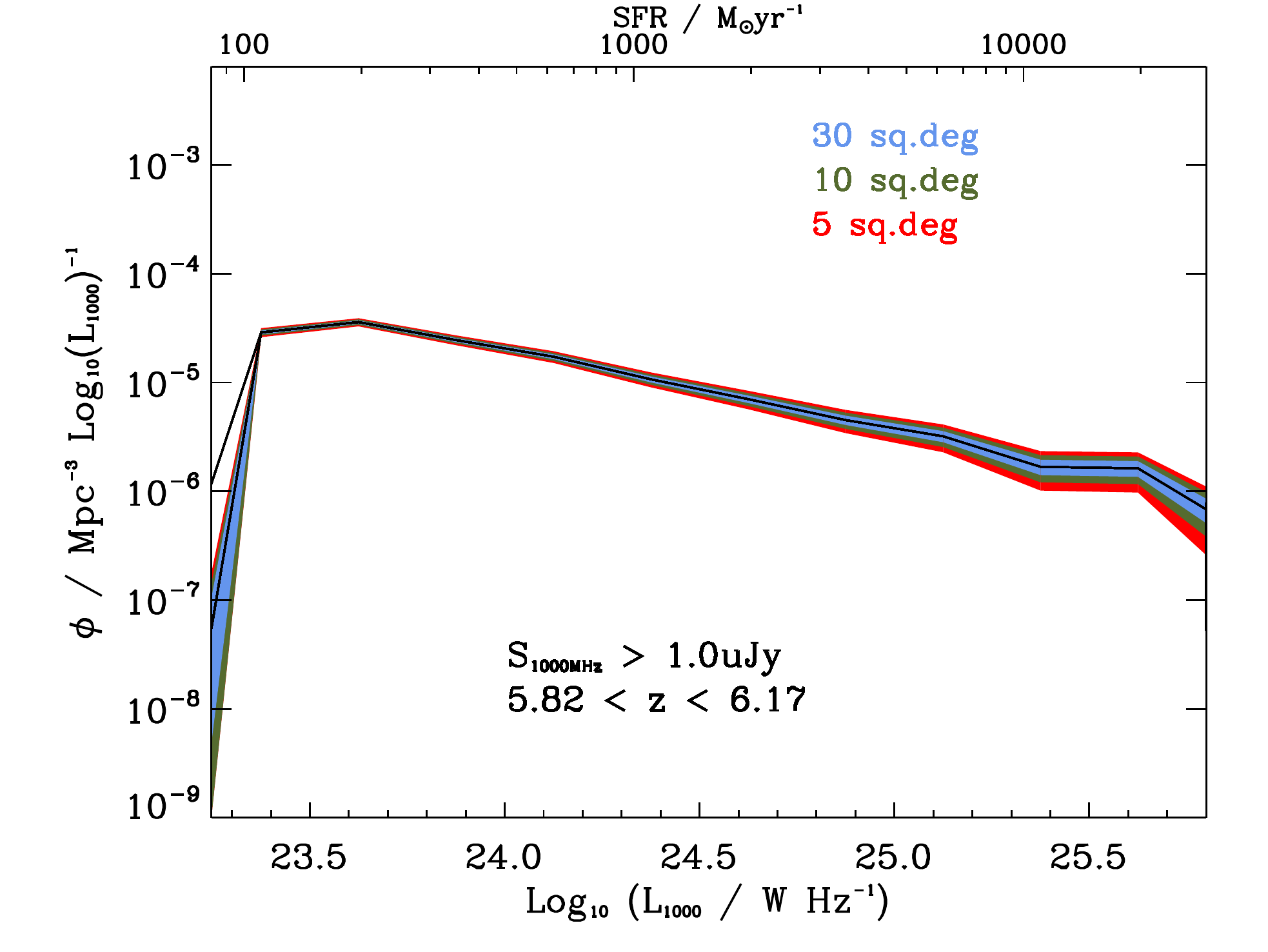}
\includegraphics[bb=40 0 566 425, scale=0.27]{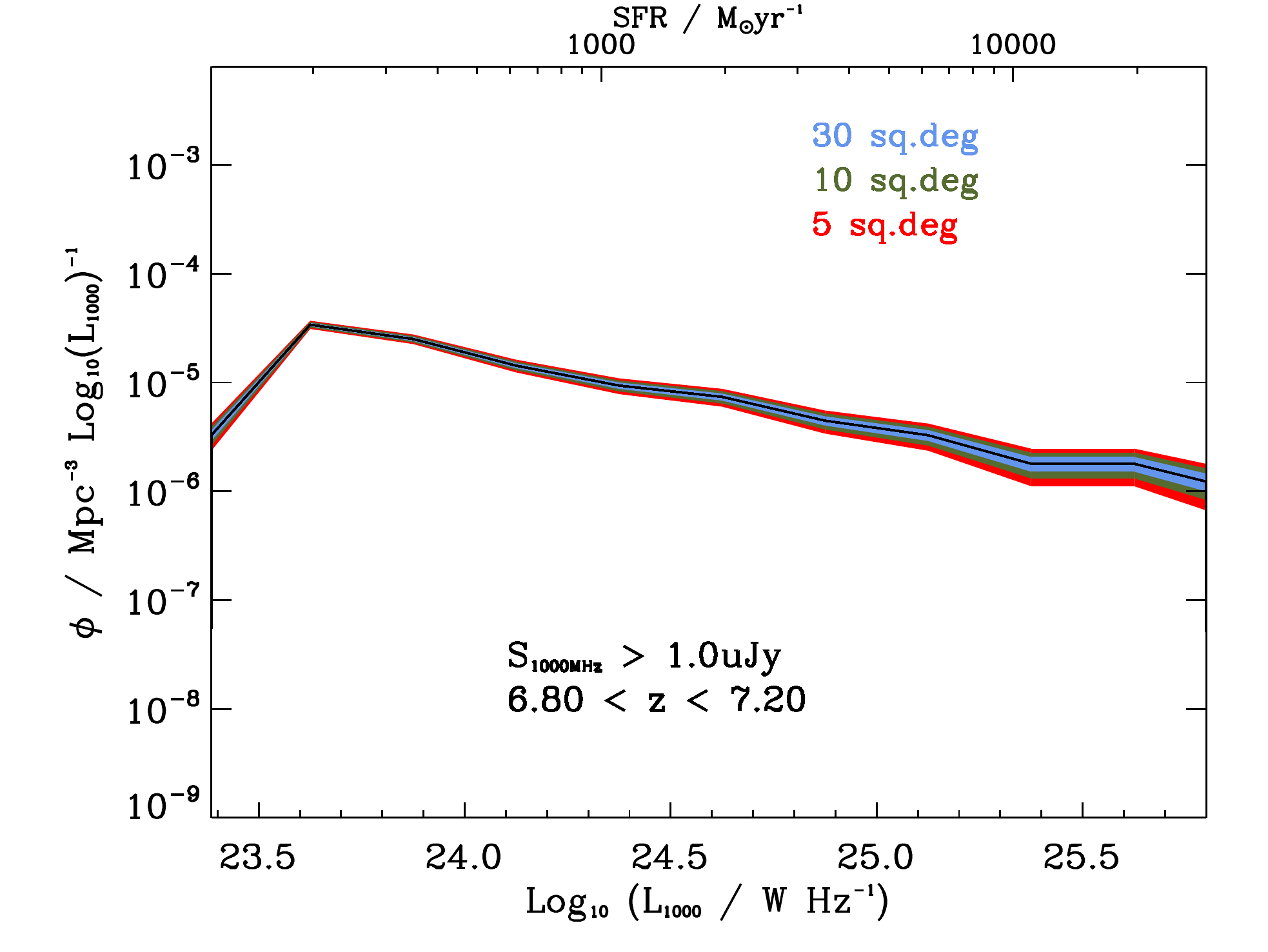}
\caption{Radio AGN luminosity functions in various redshift bins (indicated in the panels) predicted for the deep SKA1 tier based on the SKADS simulations \citep{wilman08,wilman10}. 
 \label{fig:lfsSKA2}
  }
\end{figure}

To date, the 20~cm radio luminosity function for low radio power
AGN (\lum~$\lesssim10^{25}$~\wh ) has been mostly 
derived out to $z=1.3$ \citep{sadler07,donoso09, smolcic_etal09a,simpson12,mcalpine13}. This also provided the first direct, radio-based, (albeit uncertain) observational support for radio-mode AGN feedback beyond the local universe \citep{smolcic_etal09a}. This is illustrated in Figures \ref{fig:lfs} and \ref{fig:heat}. In \f{fig:lfs} \  we show the 1.4~GHz radio luminosity functions for red, quiescent galaxies drawn from the COSMOS two\,square degree survey \citep{scoville07,ilbert10} out to $z=3$ (\smo \ et al., in prep). The luminosity functions were derived from the VLA-COSMOS 1.4~GHz Large Project \citep{schinnerer07} reaching an rms of $10(15)~\mu$Jy/beam over an area of 1(2) deg$^2$. Stacking of the red, quiescent host galaxy population, selected following \citet{ilbert10}, in the radio map constrained the luminosity function beyond $z=1.3$. The monochromatic luminosity was then converted to a kinetic power via scaling relations drawn from \citet{birzan08} and \citet{osullivan11}. 
Integrating over the  kinetic power averaged over comoving volume then yielded the heating rate exerted by  radio luminous AGN onto their surroundings as a function of cosmic time out to $z=3$, as shown in \f{fig:heat} \ \citep[see e.g.][for details]{smolcic_etal09a}. We stress that the result strongly depends on the i) knowledge of the low-luminosity end of the RL AGN luminosity function, and ii)  conversion between monochromatic radio luminosity and kinetic power. As shown in the top panel of \f{fig:heat}  \, when using the B{\^i}rzan et al.\ scaling relation to convert between monochromatic radio luminosity and kinetic power the overall heating curve systematically rises depending on the low-luminosity boundary applied to the integral. This clearly illustrates the importance of both, i) constraining the conversion between monochromatic radio luminosity and kinetic power, and ii) constraining the luminosity functions of low-power radio AGN with high precision, especially at the low-luminosity end and out to the highest redshifts possible. The first is described in more detail in  \citet{kapinska15}, and the latter, a topic of this Chapter, will be possible to resolve with the SKA1 deep and wide continuum surveys, in conjunction with deep multi-wavelength data-sets. The assumed 5$\sigma$ limits of the wide and deep SKA1 surveys correspond to a 1.4 GHz luminosity limit of $10^{23}$~\wh \  (wide) and $2\times10^{22}$~\wh \  (deep) at $z=2$, and $10^{24}$~\wh \  (wide) and $2\times10^{23}$~\wh \  (deep) at $z=6$. This will allow a direct examination of the faint end of the radio AGN luminosity function. Stacking in the radio maps will push this limit even further, while SKA2 will provide an order of magnitude push in sensitivity and determination of the faint end of the radio AGN luminosity function, as illustrated in Figs.~\ref{fig:lfsSKA1} and \ref{fig:lfsSKA2}.

\section{Studying AGN activity over cosmic time with the SKA}
SKA1 will provide a vital contribution to the science cases described above. Based on the SKADS \citep{wilman08,wilman10} simulation the SKA1 wide survey (assuming an rms of $\sim1~\mu$Jy/beam over 5000~deg$^2$ achieved in 1 year of observations) is expected to detect 
about $3\times10^7$, while the  SKA1 deep survey (assuming an rms of 0.2~$\mu$Jy/beam over 30~deg$^2$ for 2000~hours of observations) is expected to detect $4\times10^5$ AGN at 1~GHz. SKA2 will provide $10\times$ higher sensitivities at this frequency. The results discussed above can serve as a path-finder for SKA survey planning, outlined as follows:

\begin{itemize}

\item At flux densities $\lesssim 100~\mu$Jy, most sources will be SFGs and most AGN will be of the
radio-quiet type. Since the radio emission of these two classes may be powered by the same
mechanism pure continuum radio observations alone will not be able to distinguish between these populations;

\item The classification of SKA radio sources will require ancilliary,
multi-wavelength information. The location of the SKA surveys will therefore
need to be planned carefully, in conjunction with existing and planned multi-wavelength surveys; 

\item The radio band, at sensitivities that will be provided by the SKA, carries the potential of becoming the optimal band to study the evolution of the most common types 
of AGN (the radio-quiet ones) given that radio emission is simultaneously insensitive to dust and provides high angular resolution, contrary to most other wavelength regimes; 

\item The SKA1 wide and deep surveys, in conjunction with multi-wavelength data will provide the basis to resolve the long-standing quasar radio-loudness dichotomy,  unambiguously determine the source of radio-emission in RQ AGN, and study the cosmic evolution of faint radio AGN out to the highest redshifts  ($z\sim6$);

\item Radio surveys have reached such depths that they are now dominated by the same galaxies detected
by IR, optical and X-ray surveys. As a result, the SKA radio
surveys will be an increasingly important component of multi-wavelength studies of galaxy
formation and evolution and will therefore be useful to a very broad community. 

\end{itemize}

\acknowledgments{
VS acknowledges support from the European Union's Seventh Frame-work program under grant agreements 337595 and  333654. MJJ and MV acknowledge support by the South African Square Kilometre Array Project, the South African National Research Foundation. J.A. gratefully acknowledges support from the Science and Technology Foundation (FCT, Portugal) through the research grant PTDC/FIS-AST/2194/2012 and PEst-OE/FIS/UI2751/2014.
}

\bibliographystyle{apj}
\bibliography{AGN-SKA2}

\end{document}